\documentclass[reprint,twocolumn,showpacs,showkeys,preprintnumbers,amsmath,amssymb,floatfix,aps,pra,longbibliography]{revtex4-1}

\usepackage[utf8]{inputenc}
\usepackage{graphicx}% Include figure files
\usepackage{bm}% bold math

\usepackage{physics}
\usepackage{placeins}
\usepackage{graphics}
\usepackage{color}
\usepackage{hyperref}
\usepackage{multirow}
\usepackage{blindtext}

\usepackage{xcolor}
\hypersetup{
    colorlinks,
    linkcolor={red!50!black},
    citecolor={blue!50!black},
    urlcolor={blue!80!black}
}

\usepackage{pdfpages}

\makeatletter
\AtBeginDocument{\let\LS@rot\@undefined}
\makeatother

\begin{document}

%\preprint{APS/123-QED}

% \title{
% Local Variations of Proximity Exchange \\ in Cobalt/hBN/Graphene Spin Injection Geometries}

\title{Resonant magnetic proximity hot spots in Co/hBN/graphene}

\author{Klaus Zollner$^{1}$}
\email{klaus.zollner@physik.uni-regensburg.de}
\author{Lukas Cvitkovich$^{1}$}
\author{Riccardo Silvioli$^{2}$}
\author{Andreas V. Stier$^{2}$}
\author{Jaroslav Fabian$^{1,3}$}
\affiliation{$^{1}$Institute for Theoretical Physics, University of Regensburg, 93053 Regensburg, Germany}
\affiliation{$^{2}$Walter Schottky Institut and TUM School of Natural Sciences, Technische Universit\"at M\"unchen, 85748 Garching, Germany}
\affiliation{$^{3}$Halle-Berlin-Regensburg Cluster of Excellence CCE, Germany}

\begin{abstract}
 Magnetic proximity effects in $\text{Co}/\text{hBN}/\text{graphene}$ heterostructures are systematically analyzed via first-principles calculations, demonstrating a pronounced localized spatial variation of the induced spin polarization of graphene's Dirac states. The proximity-induced exchange coupling, magnetic moments, and tunneling spin polarization (TSP) are shown to depend sensitively on the atomic registry at
 the interfaces. We analyze more than twenty distinct stackings—including high- and low-symmetry configurations—and reveal that the spin splittings of graphene’s Dirac bands span a wide range from 1 to 100~meV, depending
 on the local hybridization of Co $d_{z^2}$, hBN $p_z$, and graphene $p_z$ orbitals. 
 The strongest proximity effects emerge at geometric resonances, or “proximity hot spots,” where the three orbital states overlap maximally. The local spin polarization also depends sensitively on energy: Dirac states aligned with resonant Co orbitals experience the most pronounced exchange interaction. At these energies, the pseudospin Hamiltonian description of magnetic proximity effects breaks down. Outside these resonances, the pseudospin picture is restored. Our findings highlight the intrinsically local nature of proximity effects, governed by the spectral resonance and interlayer wavefunction overlap. We further quantify how additional hBN layers, interlayer twist, and multilayer graphene modify the proximity exchange and TSP, offering microscopic insight for designing spintronic van der Waals heterostructures with engineered interfaces and optimized spin transport.
\end{abstract}

\maketitle

%------------------------------------------------------------
\section{Introduction}
%------------------------------------------------------------

Spintronics aims to utilize the spin degree of freedom of electrons in addition to their charge, enabling novel paradigms in information storage, processing, and transport with potentially lower energy consumption and higher operational speed \cite{Zutic2004:RMP, Zutic2019:MT, Han2014:NN}. Two-dimensional (2D) materials, such as graphene and transition metal dichalcogenides (TMDCs), are particularly promising for spintronic applications due to their high mobility, tunable band structures, and the ability to form van der Waals (vdW) heterostructures. However, most pristine 2D materials lack intrinsic magnetism or significant spin-orbit coupling (SOC), which are essential for active spin control. As such, an important research avenue involves engineering spin-dependent phenomena in 2D systems via proximity effects, i.e., by placing them in contact with functional materials such as ferromagnets or materials with strong SOC \cite{Avsar2019:arxiv}.

This interfacial engineering enables magnetic and spin-orbit interactions to propagate from substrates into adjacent 2D layers. The beauty of vdW engineering lies in the fact that spin interactions can be transferred without direct chemical bonding or crystal structure modification. Thus, the \textit{proximity-induced (magnetic) exchange interaction} endows graphene with magnetism, when it is placed on ferromagnetic metals through insulating barriers such as hexagonal boron nitride (hBN) \cite{Zollner2016:PRB, Leutenantsmeyer2018:PRL, Ghiasi2017:NL}. Indeed, first-principles calculations predict proximity-induced spin splittings in the graphene Dirac bands on the order of tens of meV \cite{Zollner2016:PRB}, which has been corroborated experimentally by nonlocal spin transport measurements and tunneling magnetoresistance \cite{Kamalakar2016:SR}. These proximity-induced phenomena are key to enabling efficient spin injection and manipulation in 2D materials, and they form the basis for a new generation of vdW spin valves, magnetic tunnel junctions, and spin transistors.

Among the tunable parameters of 2D heterostructures, the twist angle between adjacent layers plays a key role, giving rise to moiré patterns and allowing control over interlayer coupling and proximity effects. For instance, it has been theoretically predicted that twisting graphene on a Cr$_2$Ge$_2$Te$_6$ (CGT) substrate can reverse the sign of the proximity exchange interaction \cite{Zollner2022:arxiv}. Similarly, in TMDC/CrI$_3$ heterostructures, the valley splitting can be tuned by more than an order of magnitude via twisting, which directly modulates the orbital hybridization at the interface \cite{Zollner2023:PRB}. These findings point to the importance of local atomic arrangements in determining proximity-induced properties.

In this work, we use density functional theory to study how the proximity-induced exchange coupling varies locally in Co/hBN/graphene heterostructures. These stacks are commonly used in spin-injection devices---both in conventional lateral geometries \cite{Guimaraes2014:PRL, Gurram2017:2DM, Avsar2019:arxiv} and, more recently, in one-dimensional edge-contact setups \cite{guarochico2022tunable}---as well as in magnetic tunnel junctions \cite{dayen2020two} . Previous theory \cite{Lazic2014:PRB, Zollner2016:PRB}has shown that graphene can acquire a sizable (10 meV) exchange field from cobalt even through one or two layers of hBN, due to the alignment of Co d-states with graphene’s Dirac point. What has been missing so far is a systematic study of how this exchange effect varies across the atomic structure, and how it depends on twisting, particularly in light of the recently introduced concept of pseudospin-preserving versus pseudospin-breaking proximity effects (see Ref. \cite{cvitkovich2025machinelearningpredictionmagnetic}), demonstrated in graphene/Cr$_2$Ge$_2$Te$_6$ slabs by some of the present authors.

We analyze more than twenty high- and low-symmetry Co/hBN/graphene stacking configurations to quantify the impact of local atomic registries on Dirac band spin splittings, induced magnetic moments, and tunneling spin polarization (TSP). Our results reveal that even nominally uniform heterostructures may exhibit spatial inhomogeneity in spin injection and transport. These effects can be understood as emerging from spatially varying hybridization across the moiré unit cell, leading to position-dependent spin splittings and anisotropies.
In particular, we demonstrate that the induced spin splittings in graphene can vary from 1 to 100 meV, with local magnetic moments and tunneling spin polarizations (TSPs) also displaying strong spatial dependence. Regarding TSPs, specific local geometries can give rise to spin-filtering "hot spots" in realistic device structures.

A key insight revealed in our study is that proximity effects are inherently local. Because they depend on orbital hybridization, particularly between out-of-plane orbitals (graphene \( p_z \), N \( p_z \), and Co \( d_{z^2} \)), the atomic registry and relative alignment between layers become critically important. Even small variations in stacking - induced by lattice mismatch or intentional twist - can lead to dramatic modulations of the proximity-induced interactions.
The degree of this hybridization, and therefore the strength and character of the proximity effect, depends sensitively on the exact stacking configuration and interlayer distances.
Our results have direct implications for the design of spintronic devices and suggest new strategies for tuning interfacial magnetic interactions at the atomic scale.

The manuscript is organized as follows. In Section \ref{Sec:Structural_Setup}, we introduce the structural setup of the graphene/hBN/Co heterostructures that we consider in our first-principles calculations. In Section \ref{Sec:Computational_Details}, we summarize the DFT calculation details we employ to unveil different band-structure effects and tunneling conductances, based on different local atomic stacking configurations. Exemplary input files are listed in the Supplemental Material, see~\footnotemark[1]. In Section \ref{Sec:Local_Proximity_Exchange}, we compare band structure and density of states of selected stacking configurations and highlight the impact of the local atomic registries on proximity physics. These results are mapped to a more spatially extended supercell of graphene/hBN/Co. Section \ref{Sec:Local_TDOS} provides insights into the relationship between tunneling spin polarization and the associated local atomic stacking. 
In Section \ref{Sec:Atomic_Site_Overlap}, we analyze the overlap of atomic orbitals in three-layer atomic structures as a function of the twist angle using geometric modeling. 
Finally, in section \ref{Sec:Summary} we summarize the   main findings of our research and conclude the manuscript.

%------------------------------------------------------------
\section{Structural Setup}
%------------------------------------------------------------
\label{Sec:Structural_Setup}

 The graphene/hBN/Co heterostructures are set up with the {\tt atomic simulation environment (ASE)} \cite{ASE} and the {\tt CellMatch} code \cite{Lazic2015:CPC}, implementing the coincidence lattice method \cite{Koda2016:JPCC,Carr2020:NRM}. We consider small, medium-sized, and
 large-scale atomic structures, to illustrate the effects of local atomix registries on the magnetic proximity effects. 

\begin{figure*}[!htb]
	\includegraphics[width=0.8\textwidth]{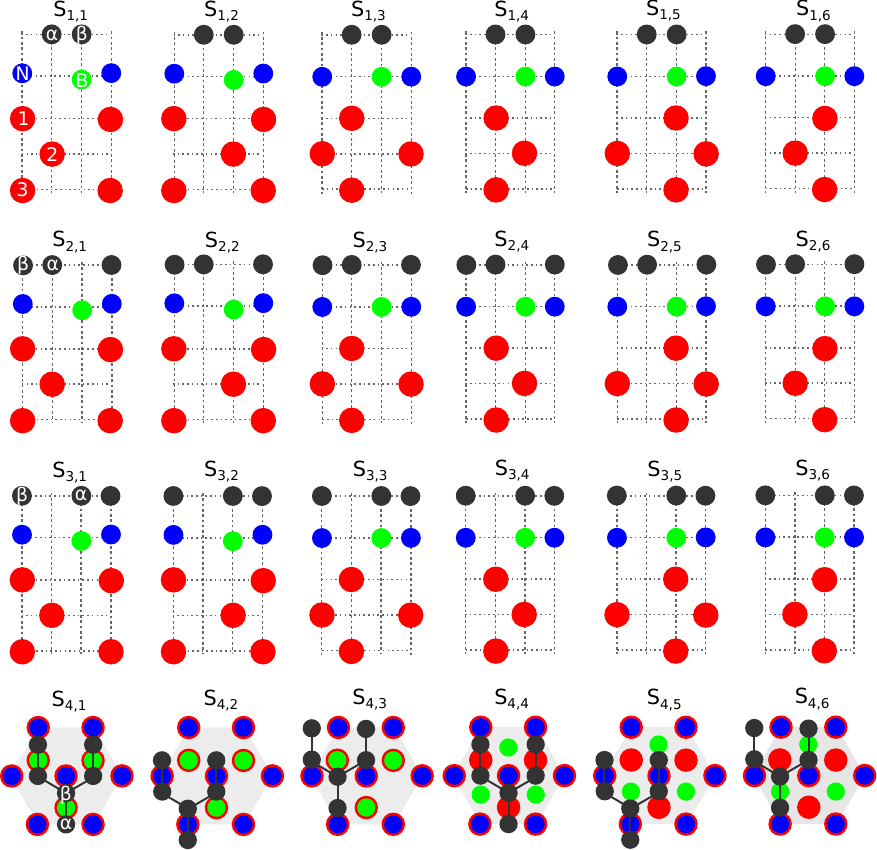}
	\caption{Sketch of the geometries, showing the different stacking configurations S$_{1,1}$ -- S$_{3,6}$, of the 18 lattice-matched high-symmetry commensurate structures. The colors of the spheres correspond to different atoms (black = C, green = B, blue = N, red = Co). Within each row, the graphene/hBN stacking is fixed. Within each column, the hBN/Co stacking is fixed. Labels $\alpha$ and $\beta$ represent the graphene sublattices, while labels $1-3$ represent the different Co layers. The rippling of hBN, if present, is indicated. Structures S$_{4,1}$ -- S$_{4,6}$ in the last row have lowered symmetry and are derived from geometries S$_{1,2}$ and S$_{1,1}$.    
    Further details are summarized in Table~\ref{Tab:Structures}.
 \label{Fig:Structure}}
\end{figure*}

Small structures are shown in Fig.~\ref{Fig:Structure}.
In total, we consider 18 high-symmetry lattice-matched commensurate stacks $S$, and six additional stackings---derivatives of geometries S$_{1,1}$ and S$_{1,2}$---with reduced symmetry that may appear in moir\'e superlattices. The lattice constant of graphene is $a=2.46$~${\textrm{\AA}}$ \cite{Neto2009:RMP}, the one of hBN is $ a=2.504$~${\textrm{\AA}}$ \cite{Catellani1987:PRB}, and the one of hcp-cobalt is $a=2.507$~${\textrm{\AA}}$ \cite{Singal1977:PRB}. Here, we fix an effective average lattice constant of $a=2.4903$~${\textrm{\AA}}$ for this well lattice-matched system, as a compromise to make the lattices commensurable and to keep the unit cells small to be considered as representative atomic registries which can be identified within larger systems. The corresponding strain is about $\pm 1.2$\% for the individual material components. 

 %------------------------------------------------------------------------
%------------------------------------------------------------------------

  In all structures, the individual monolayers are barely strained, so the extracted band offsets and magnetic proximity effects are representative when compared to those in experiment. To simulate quasi-2D systems, we add a vacuum of about $20$~\AA~to avoid interactions between periodic images in our slab geometries. 

%------------------------------------------------------------
\section{Computational Details}
%------------------------------------------------------------
\label{Sec:Computational_Details}
For reproducibility, we provide representative input files for all calculations done on the S$_{1,2}$ configuration, in the Supplemental Material~\footnotemark[1]. 

\subsection{DFT calculations}
The electronic structure calculations and structural relaxations of the graphene/hBN/Co heterostructures are performed by DFT~\cite{Hohenberg1964:PRB} 
with {\tt Quantum ESPRESSO v7.2}~\cite{Giannozzi2009:JPCM,QE-2017,QE-exa}. Self-consistent calculations are carried out with a $k$-point sampling of $300\times 300\times 1$. 
We use an energy cutoff for charge density of $800$~Ry and the kinetic energy cutoff for wavefunctions is $75$~Ry for the scalar relativistic pseudopotentials
with the projector augmented wave method~\cite{Kresse1999:PRB} with the 
Perdew-Burke-Ernzerhof exchange correlation functional~\cite{Perdew1996:PRL}.
Open shell calculations provide the spin polarized ground state.
For the self-consistent calculation, we employ a threshold of $1\times10^{-8}$~Ry and Fermi-Dirac smearing of $5\times10^{-4}$~Ry.
For the relaxation of the heterostructures, we add DFT-D2 vdW corrections~\cite{Grimme2006:JCC,Grimme2010:JCP,Barone2009:JCC} and use 
quasi-Newton algorithm based on trust radius procedure. 
To get proper interlayer distances and to capture possible moir\'{e} reconstructions, we allow all atoms to move freely within the heterostructure geometry during relaxation. Relaxation on the high-symmetry structures is performed until every component of each force is reduced below $1\times10^{-4}$~Ry/$a_0$, where $a_0$ is the Bohr radius. In the case of reduced-symmetry stackings, the atoms are only allowed to relax in $z$-direction, otherwise the atoms would rearrange to a high-symmetry stacking. Spin-orbit coupling is omitted in the calculations, as it shows negligible impact compared to the exchange coupling, justified on exemplary calculation results shown in SM~\footnotemark[1].
The density of states (DOS) is calculated with the optimized tetrahedron method~\cite{Kawamura2014:PRB} and magnetic moments are calculated via Löwdin population analysis~\cite{Lowdin1950:JCP} as implemented in {\tt Quantum ESPRESSO}.
For each structure, we calculate the total energy, average interlayer distances, induced magnetic moments, and the rippling of the hBN layer (deviation of B/N $z$-positions from the average). Note that the hBN rippling is always such that the B atom is closer to the Co interface than the N atom.

%------------------------------------------------------------------------
\begin{figure*}[!htb]
	\includegraphics[width=0.85\textwidth]{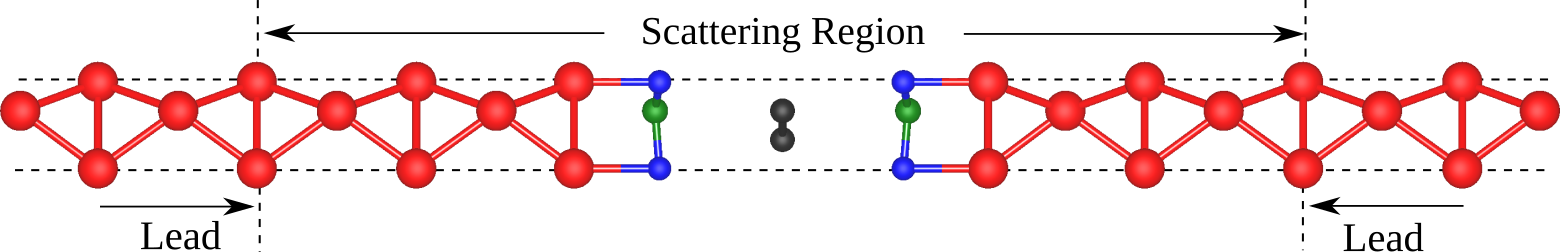}
	\caption{Setup for the transmission calculations. For the transmission calculations with {\tt PWCOND}, we consider a Co/hBN/graphene/hBN/Co stack as the scattering region, with semi-infinite bulk Co leads.
 \label{Fig:Tunneling}}
\end{figure*}
%------------------------------------------------------------------------

\subsection{Transmission calculations}
The transmission calculations are performed by {\tt PWCOND} implemented in {\tt Quantum ESPRESSO} package \cite{QE-2017}. By considering a lead/scattering-region/lead setup, the code solves the quantum mechanical scattering problem and calculates the ballistic conductance. We consider symmetric tunneling geometries schematically shown in Fig.~\ref{Fig:Tunneling}, which are based on the stacking configurations in Fig.~\ref{Fig:Structure}.  The atomic positions of the scattering regions are fully relaxed prior to transmission calculations. The magnetizations of the Co leads are parallel, as we are interested in the variation of the spin-dependent transmission across different atomic registries. 

When performing the self-consistent calculation for the scattering region (leads), we employ the same parameters and pseudopotentials as above, but a $k$-point grid of $60\times 60\times 1$ ($36\times 36\times 24$). 
The specific input for {\tt PWCOND} is given in the Supplemental Material~\footnotemark[1].  
From the spin-dependent transmissions, $T_{\uparrow / \downarrow}$, we extract the tunneling spin polarization (TSP) as:
\begin{equation}
    P_T = \frac{T_{\uparrow}-T_{\downarrow}}{T_{\uparrow}+T_{\downarrow}}.
\end{equation}
This polarization is a function of the electron energy and, as we show below, depends strongly on the atomic registries. 

%------------------------------------------------------------
\section{Local Variations of Proximity Exchange}
%------------------------------------------------------------
\label{Sec:Local_Proximity_Exchange}

%------------------------------------------------------------------------
\begin{figure*}[!htb]
	\includegraphics[width=0.9\textwidth]{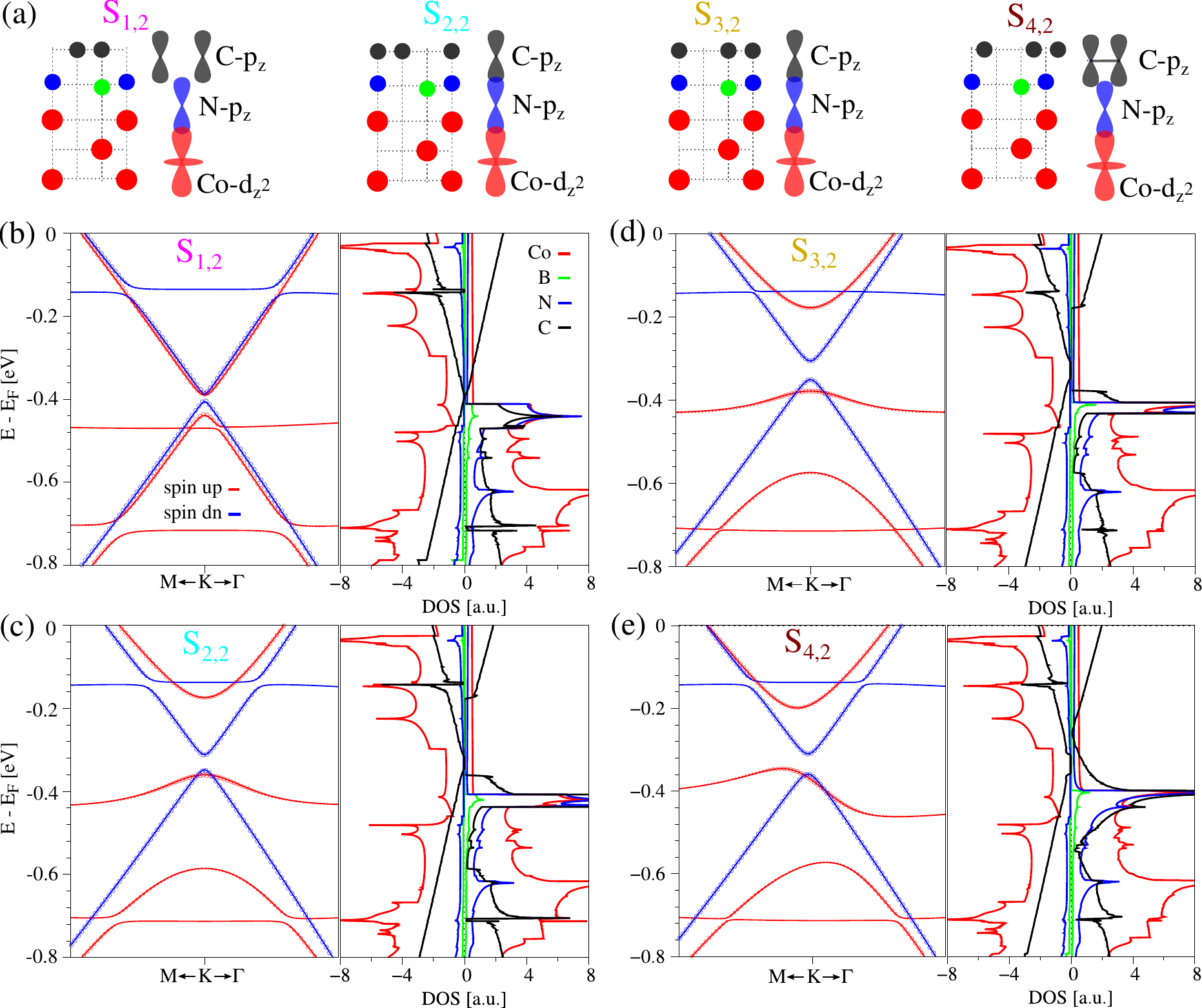}
	\caption{(a) Side views of exemplary high-symmetry stacking configurations S$_{1,2}$, S$_{2,2}$, S$_{3,2}$, and S$_{4,2}$. The hybridization channels from Co $d_{z^2}$ to C $p_z$ orbitals via N $p_z$ orbitals are sketched. (b) Zoom to the proximitized low energy Dirac bands ($E_F = 0$) and the corresponding spin- and atom-resolved density of states of the S$_{1,2}$ configuration. The open circles on the bands represent the projection on graphene orbitals. Positive (negative) DOS corresponds to majority (minority) spin channels. (c,d,e) The same as (b), but for the corresponding stacking configuration as labeled in the dispersion. \label{Fig:S12_S22_bands_DOS_geometries}}
\end{figure*}
%------------------------------------------------------------------------

%------------------------------------------------------------------------
\begin{figure*}[!htb]
\centering
	\includegraphics[width=0.9\textwidth]{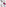}
	\caption{ (a) Medium-sized Co/hBN/graphene supercell. The different colored spheres emphasize the different local high-symmetry stackings. The dashed line represents the unit cell. (b) The proximity-induced calculated magnetic moments on C atoms, overlayed on the geometry. (c) DFT-calculated spin-resolved band structure in the vicinity of the K point. The open circles on the bands represent projections onto graphene orbitals. (d) DFT-calculated spin polarization. We cut through the $p_z$-orbital polarization slightly above the C atoms. Yellow/Red corresponds to positive polarization, while Cyan/Blue corresponds to negative polarization, in line with the magnetic moments shown in (b). (e) Spin polarization, taking into account states below the Dirac point at $E-E_F =-0.4 \textrm{eV}$ and energy window of $\pm 10$meV. The polarization is positive, highly non-uniform, and mainly localized around the S$_{3,2}$ configuration. (f) Same as (e), but above the Dirac point at $E-E_F =-0.2 \textrm{eV}$. The polarization is negative and uniformly distributed.
 \label{Fig:Medium_bands_moments_LDOS}}
\end{figure*}
%------------------------------------------------------------------------

We wish to convey the idea that the precise atomic stacking registry has a significant impact on both proximity-induced magnetic exchange coupling, which modifies the Dirac bands of graphene, and on the tunneling spin-injection efficiency.
The following analysis should be generally applicable to vdW interfaces. 

%-----------------------------------------------------------------
\begin{table*}[htb]
\begin{ruledtabular}
\caption{\label{Tab:Structures} Total energies with respect to the lowest energy reference structure S$_{1,2}$, averaged interlayer distances between the layers, d$_{\textrm{Co/hBN}}$ and d$_{\textrm{hBN/Gr}}$, rippling of the hBN layer, $\delta_{z}$, and the induced atomic magnetic moments on boron (B), nitrogen (N), and carbon atoms at $\alpha$ and $\beta$ sublattice sites in graphene.}
\begin{tabular}{lcccccccc}
config. & $E_{\textrm{tot}}-E_{0}$ [meV]  & d$_{\textrm{Co/hBN}}$ [\AA] & $\delta_{z}$  [\AA] & d$_{\textrm{hBN/Gr}}$ [\AA]  & B [$10^{-3} \mu_B$] & N [$10^{-3} \mu_B$] & $\alpha$ [$10^{-3} \mu_B$] & $\beta$ [$10^{-3} \mu_B$]    \\ \hline
S$_{1,1}$ & 4.305 & 2.040 & 0.059 & 3.071 & -41.99 & 15.38 & 0.43 & 0.25 \\
S$_{1,2}$ & 0 & 2.039 & 0.060 & 3.072 & -45.14 & 13.15 & 0.36 & 0.31  \\
S$_{1,3}$ & 281.824 & 2.992 & 0.007 & 3.121 & -1.92 & 0.79 & 0.08 & -0.18 \\
S$_{1,4}$ & 282.429 & 2.994 & 0.007 & 3.120 & -2.07 & 0.12 & 0.02 & -0.14 \\
S$_{1,5}$  & 300.364 & 3.102 & 0.003 & 3.122 & -1.83 & 5.51 & -0.25 & 0.18 \\
S$_{1,6}$  & 302.107 & 3.107 & 0.003 & 3.122 & -1.33 & 4.78 & -0.25 & 0.16 \\
S$_{2,1}$  & 30.700 & 2.038 & 0.060 & 3.210 & -42.38 & 14.79 & 3.89 & -3.33 \\
S$_{2,2}$  & 26.175 & 2.035 & 0.061 & 3.208 & -45.48 & 12.51 & 3.81 & -3.20 \\
S$_{2,3}$  & 310.869 & 2.994 & 0.007 & 3.301 & -1.90 & 0.48 & -0.09 & 0.08 \\
S$_{2,4}$  & 311.467 & 2.995 & 0.008 & 3.299 & -2.11 & -0.14 & -0.09 & 0.07 \\
S$_{2,5}$  & 328.462 & 3.104 & 0.004 & 3.298 & -1.95 & 5.51 & -0.12 & 0.05 \\
S$_{2,6}$  & 330.295 & 3.109 & 0.003 & 3.295 & -1.44 & 4.77 & -0.12 & 0.04 \\
S$_{3,1}$  & 32.916 & 2.038 & 0.060 & 3.236 & -43.18 & 16.23 & 3.37 & -3.45 \\
S$_{3,2}$  & 28.804 & 2.035 & 0.061 & 3.239 & -46.41 & 13.95 & 3.23 & -3.43 \\
S$_{3,3}$  & 315.285 & 2.992 & 0.008 & 3.354 & -2.09 & 0.60 & -0.18 & 0.08 \\
S$_{3,4}$  & 316.028 & 2.993 & 0.008 & 3.355 & -2.25 & -0.08 & -0.15 & 0.04 \\
S$_{3,5}$  & 334.027 & 3.103 & 0.004 & 3.359 & -1.78 & 5.42 & 0.03 & -0.02 \\
S$_{3,6}$  & 335.895 & 3.109 & 0.003 & 3.358 & -1.25 & 4.62 & 0.02 & -0.02 \\
S$_{4,1}$  & 17.419 & 2.036 & 0.060 & 3.163 & -45.87 & 13.50 & 1.35 & -1.09 \\
S$_{4,2}$  & 25.126 & 2.036 & 0.061 & 3.208 & -45.92 & 13.31 & 0.11 & 0.14 \\
S$_{4,3}$  & 16.215 & 2.036 & 0.060 & 3.152 & -45.52 & 12.96 & 1.48 & -0.97 \\
S$_{4,4}$  & 21.632 & 2.039 & 0.060 & 3.163 & -42.68 & 15.75 & 1.42 & -1.10 \\
S$_{4,5}$  & 29.439 & 2.038 & 0.060 & 3.208 & -42.75 & 15.59 & 0.13 & 0.14 \\
S$_{4,6}$  & 20.587 & 2.039 & 0.060 & 3.152 & -42.37 & 15.22 & 1.50 & -0.99 \\
\end{tabular}
\end{ruledtabular}
\end{table*}
%-----------------------------------------------------------------

%------------------------------------------------------------
\subsection{Monolayers of Graphene and hBN}
%------------------------------------------------------------
We start with the simplest scenario, considering monolayers of graphene and hBN. As shown above, we can already form 18 different high-symmetry stacking configurations when considering lattice-matched constituents. Depending on the precise atomic stacking registry, strong variations in interlayer distance and induced magnetic moments occur, see Table~\ref{Tab:Structures}. 
Note, that the magnetization of bulk Co atoms is 1.69$\mu_{\textrm{B}}$ and the spin polarization at the Fermi level is $P_N= \frac{N_{\uparrow}-N_{\downarrow}}{N_{\uparrow}+N_{\downarrow}}$ = -0.556.

The lowest energy configuration is S$_{1,2}$, with interlayer distances of about 2.0~\AA, between Co and hBN, and about 3.1~\AA, between hBN and graphene. Additionally, the hBN is rippled by about 0.06~\AA, in agreement with Ref.~\cite{Zollner2016:PRB}.
Depending on the stacking registry, interlayer distances between Co and hBN vary in the range of 2.03 -- 3.11~\AA, while the interlayer distance between hBN and graphene varies between 3.07 -- 3.36~\AA. Significant rippling of the hBN layer occurs only when the N atom is directly above the top Co atom. 
From Table~\ref{Tab:Structures}, it is also evident that the proximity-induced magnetic moments are strongly dependent on the registry. 

For all 24 registries in Fig.~\ref{Fig:Structure}, we have calculated the proximitized low-energy Dirac bands and the corresponding spin and atom-resolved DOS, see SM~\footnotemark[1]. 
Representative low-energy Dirac bands, for the S$_{1,2}$, S$_{2,2}$, S$_{3,2}$, and S$_{4,2}$ stackings, are shown in Fig.~\ref{Fig:S12_S22_bands_DOS_geometries}. We find that the registry dictates the overlap of the wavefunctions and thereby the hybridization of C $p_z$ orbitals with Co $d_{z^2}$ orbitals, mediated by N $p_z$ orbitals, as sketched in Fig.~\ref{Fig:S12_S22_bands_DOS_geometries}(a). 
Already, the four selected dispersions demonstrate that the Dirac bands can experience vastly different spin splittings, band alignments, and even anisotropies. For example, the S$_{1,2}$ configuration shows very well preserved Dirac bands, moderate proximity-induced exchange splittings, as well as moderate anticrossings with the flat bands originating from Co $d$-orbitals. In contrast, for the S$_{2,2}$ configuration, in particular the spin-up Dirac cone is not recognizable anymore as strong hybridization occurs. The reason is the direct hybridization channel, as sketched in Fig.~\ref{Fig:S12_S22_bands_DOS_geometries}(a). 
\emph{Across the stacking configurations we studied, the proximity-induced exchange spin splitting in the Dirac cone ranges from 1 to 100$~\mathrm{meV}$, with systematically larger splittings observed for holes}. As already mentioned, a key factor in the hybridization is the position of the N atom, as we explicitly demonstrate by calculating the integrated local DOS for two stacking configurations~\footnotemark[1].
Also the energetic position of the Dirac point is quite different for the stackings shown in Fig.~\ref{Fig:S12_S22_bands_DOS_geometries}. Overall, we find the Dirac point in the range of 0.2 -- 0.4~eV below the individual heterostructure Fermi level, indicating the possibility of strong doping variations across the configurations. 
When low-symmetry registries are considered, such as S$_{4,1}$, we additionally find a strong anisotropy of the low-energy Dirac band spin splittings. This is demonstrated in Fig.~\ref{Fig:S12_S22_bands_DOS_geometries}(e) and in the SM~\footnotemark[1], where we calculate the low-energy dispersion along two different Brillouin zone paths. 

When considering more realistic experimental heterostructure setups, strain is minimized, twist angles can be arbitrary, and many local stacking configurations may occur, due to the moir\'e structure from the lattice mismatch. For simplicity, we consider lattice-matched hBN and Co, and place graphene on top at  different twist angles, minimizing strain. One of such possible geometries is shown in Fig.~\ref{Fig:Medium_bands_moments_LDOS}. This supercell has 191 atoms, built by considering an aligned energetically favorable Co/hBN interface from the S$_{1,2}$ geometry of Fig.~\ref{Fig:Structure}, but with a lattice constant of 2.5051~${\textrm{\AA}}$. The graphene layer is placed at a twist angle of 10.89° above the hBN/Co interface and has a lattice constant of 2.46~${\textrm{\AA}}$. The strains for Co and hBN are -0.076\% and 0.044\%. This medium-sized structure is still computationally manageable, but the local stacking registries are less comparable to those in small lattice-matched cells due to the rather large twist angle involved. 
For the medium-sized structure, we have calculated the low-energy Dirac dispersion, DOS, and the local magnetic moments on the C atoms, see Fig.~\ref{Fig:Medium_bands_moments_LDOS} and SM~\footnotemark[1].

The low-energy Dirac bands for this geometry, see Fig.~\ref{Fig:Medium_bands_moments_LDOS}(c), are almost gapless, in contrast to those Dirac bands from the individual stackings. This results from an averaging effect of the graphene sublattice asymmetry on the various stacking configurations. Furthermore, a rather uniform proximity exchange splitting of about 35~meV arises near the Dirac point, again due to the averaging effect.

The induced magnetic moments in carbon in this supercell vary strongly within $\pm1\times10^{-3}\mu_{\textrm{B}}$, see Fig.~\ref{Fig:Medium_bands_moments_LDOS}(b). 
In fact, starting at the S$_{2,2}$ configuration, magnetic moments on C atoms are largest and opposite on A and B sublattice. Moving away towards the S$_{1,2}$ configuration, the polarizations gradually decrease. Furthermore, the sublattice polarizations even reverse sign.
Finally, in $S_{4,2}$ and $S_{4,3}$ the sublattice arrangements are very different. This variety nicely demonstrates the very local character of proximity exchange coupling and the influence of the atomic registry on spin properties.

Even though the low-energy Dirac dispersion resembles a uniform proximity-induced exchange coupling, see Fig.~\ref{Fig:Medium_bands_moments_LDOS}(c), the local nature of the coupling is clearly visible in the magnetic moments. To further support this picture, we have also calculated the local density of states (LDOS), at different energies, see SM~\footnotemark[1]. From the LDOS map at the Fermi level, we find a rather uniform charge density distribution on graphene. At lower energy, $E - E_F = -0.4$~eV, roughly near the strong anticrossing in the spin-up channel of the dispersion, the charge density concentrates around regions where C and N atoms are vertically stacked. 
This picture is in line with the dispersions of the high-symmetry stackings, i. e., the coupling between layers, and the associated band anticrossing, is concentrated to specific regions in real space. 

In Fig.~\ref{Fig:Medium_bands_moments_LDOS}(d), we also show the DFT-calculated spin polarization, which is in line with the magnetic moments. The spin polarization is calculated as the difference between the spin-up and spin-down electron densities, integrated up to the Fermi level.  In contrast, in Fig.~\ref{Fig:Medium_bands_moments_LDOS}(e,f), we show spectrally resolved local spin polarization, taking into account the states at $\pm 100$~meV above or below the Dirac point. Below the Dirac point near the anticrossing, the spin polarization is positive, 
highly non-uniform, and mainly localized around the S$_{3,2}$ configuration, similar to the LDOS. Such regions form \emph{proximity hot spots}. Above the Dirac point, the polarization is negative and uniformly distributed across graphene.  This demonstrates the dependence of the local nature of proximity physics on energy. 

We even generated a much larger supercell (1619 atoms), see SM~\footnotemark[1], which further serves to illustrate the spatial variation of magnetic proximity effects arising from different local registries $S$.

%------------------------------------------------------------
\subsection{Additional Graphene and hBN Layers}
%-----------------------------------------------------------

In experiments, often more than one monolayers of the same material are considered when building heterostructures. The results for bilayer graphene and bilayer hBN are summarized in the SM~\footnotemark[1]. Here, we briefly discuss the main findings.

With two hBN layers, the proximity coupling is strongly suppressed, as evident from the proximity-induced magnetic moments on C atoms and the low-energy Dirac band splittings. Nevertheless, a pronounced coupling between Co and C orbitals can still arise, which is now mediated by a consecutive coupling via N orbitals from the first hBN layer and B orbitals from the second hBN layer. In particular, the low energy dispersions show that band topologies are still strongly affected by the local stacking configuration. More precisely, spin splittings vary in sign and can still range up to tens of meV for specific atomic configurations.

\begin{figure}[!htb]
	\includegraphics[width=0.99\columnwidth]{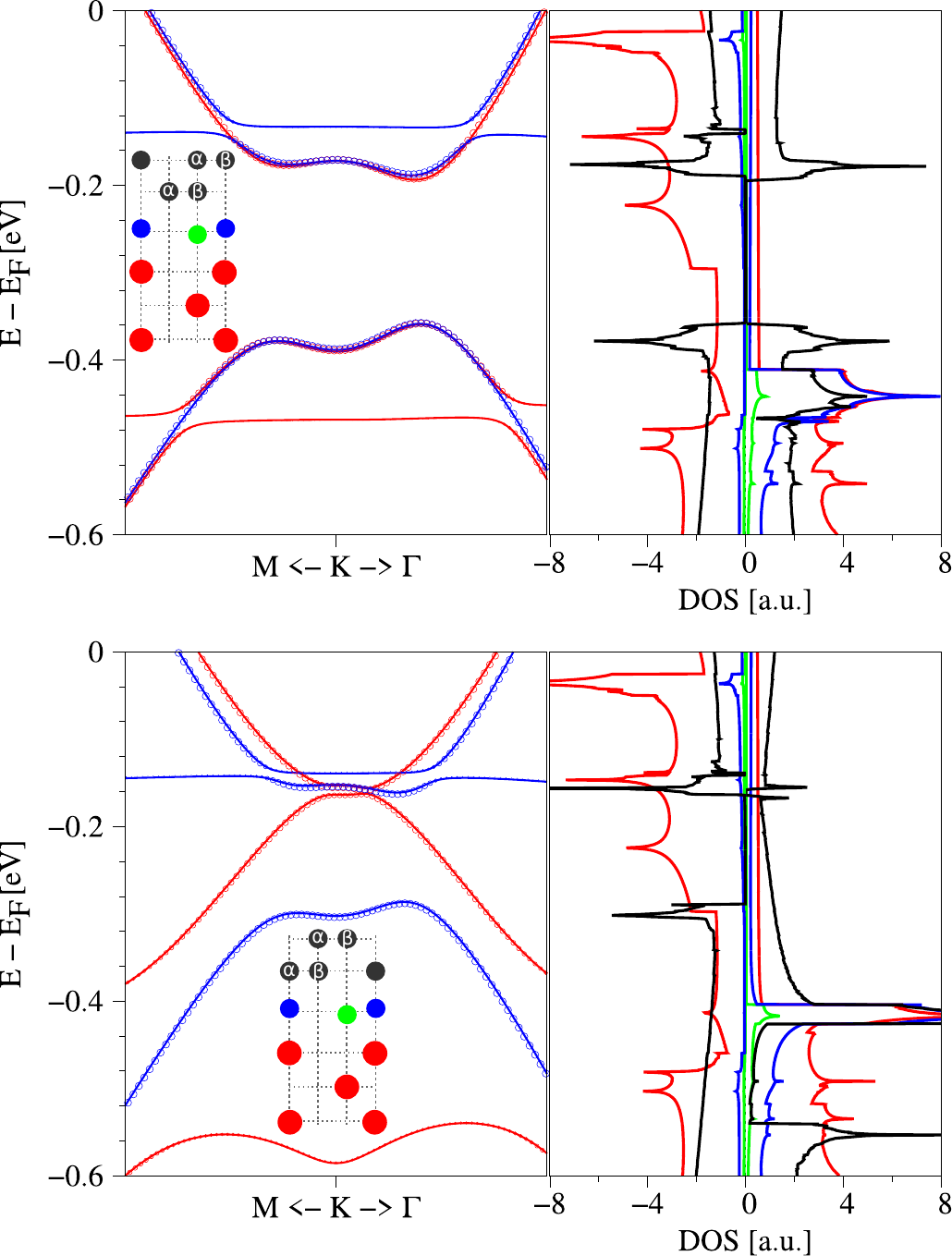}
	\caption{Zoom to the proximitized low energy bilayer-graphene bands ($E_F = 0$) and the corresponding spin- and atom-resolved density of states of selected configurations of bilayer-graphene/hBN/Co. The open circles on the bands represent the projection on bilayer-graphene orbitals. Positive (negative) DOS corresponds to the majority (minority) spin channels. The stackings are sketched in the inset.
 \label{Fig:bands_BLG_main}}
\end{figure}

In the case of bilayer graphene, the short-rangeness of proximity effect is at play. While the first graphene layer still shows pronounced proximity-induced magnetic moments, up to about $\pm 4 \times 10^{-3} \mu_{\textrm{B}}$ for some stackings and in line with the monolayer graphene results, the magnetic moments in the second graphene layer are further suppressed by one order of magnitude.
The band gap in the bilayer graphene spectrum is quite large, on the order of 200~meV for almost all cases, see Fig.~\ref{Fig:bands_BLG_main}. Certainly, the interplay of short-ranged proximity-exchange and the layer degree-of-freedom is also recognizable, given that the bilayer graphene conduction and valence bands show rather different spin splitting strengths~\cite{Zollner2018:NJP}.
However, we also find one rather special case, in which the spin-down channel shows a gapped spectrum, while the spin-up channel is nearly gapless, see Fig.~\ref{Fig:bands_BLG_main}. Hence, this local stacking configuration would support 100\% polarized islands within otherwise insulating stackings inside moiré structures. Eventually, one can think of spin-polarized network channels within the bilayer graphene.

%------------------------------------------------------------------------
%------------------------------------------------------------------------

%------------------------------------------------------------
\section{Local Variations of Tunneling Spin Polarization}
%------------------------------------------------------------
\label{Sec:Local_TDOS}
In Ref. \cite{Faleev2015:PRB}, it was demonstrated that Co/hBN interfaces show a Brillouin zone spin filtering mechanism due to the specific complex band structure of hBN. 
Additionally, in Refs.~\cite{Karpan2011:PRB, Karpan2007:PRL} it was demonstrated that graphene acts as an ideal spin filter supporting minority carriers of hcp-Co. 
The reason is that near the Fermi level, graphene Dirac states are present only near K, where also only minority carriers of Co are present. 

In our case, since each different stacking registry provides vastly different results in terms of low-energy Dirac dispersion, DOS, and magnetic moments, we are additionally interested in the local tunneling spin polarization (TSP). We believe that in experimental setups, there can be \emph{hot spots}, where spins are most efficiently injected, based on the ideal wavefunction overlap across the vdW interface, and spin filtering effects~\cite{Karpan2007:PRL,Karpan2011:PRB,Faleev2015:PRB}. Another factor is certainly the band alignment of the different configurations, leading to inhomogeneity in spin injection. 

We consider the definition of the TSP in Sec.~\ref{Sec:Computational_Details}. 
For six representative structures,  S$_{1,2}$, S$_{2,2}$, S$_{3,2}$, S$_{4,1}$, S$_{4,2}$, and S$_{4,3}$ we calculate the TSP. This selection is motivated by the appearance of these atomic registries in the medium and large heterostructures we consider. The calculation setup for the TSP is shown in Fig.~\ref{Fig:Tunneling}.

\begin{figure}[!htb]
	\includegraphics[width=0.99\columnwidth]{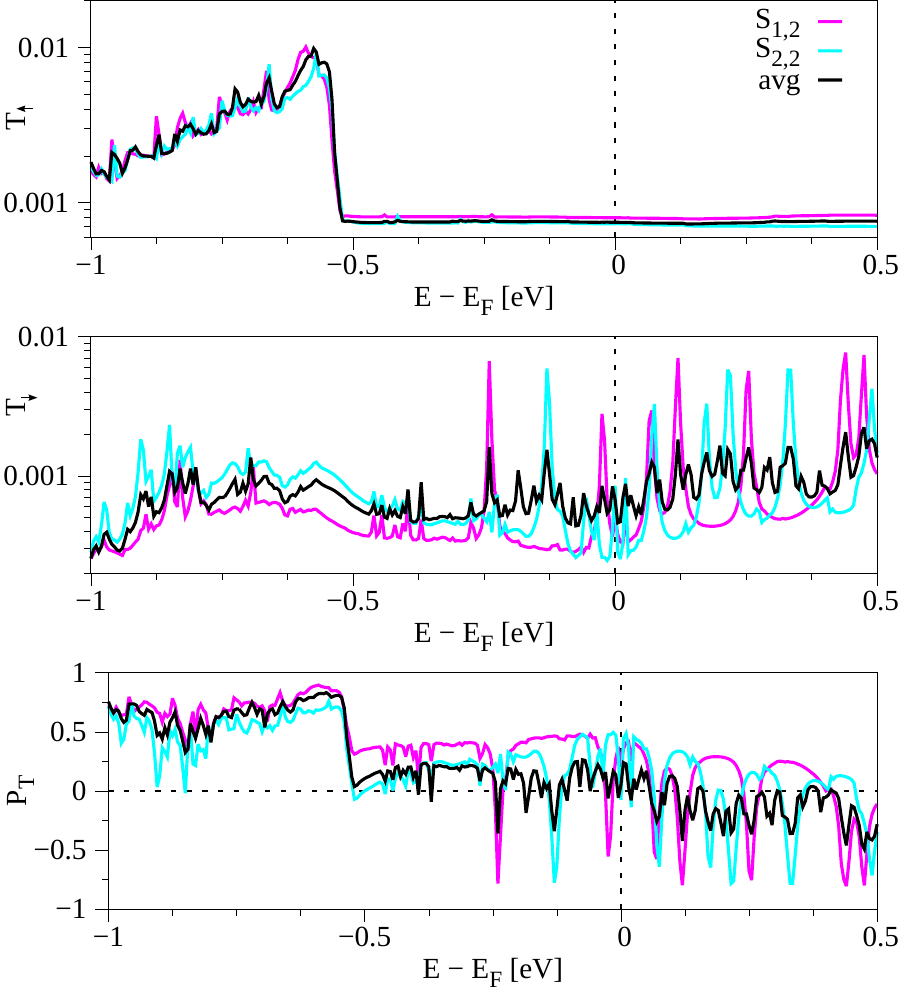}
	\caption{The spin-resolved transmissions, $T_{\uparrow / \downarrow}$, and the tunneling spin polarization, $P_T$. We present results for two exemplary stacking configurations, as well as for the average when considering the 6 configurations from Fig.~\ref{Fig:Medium_bands_moments_LDOS}(a). Results for all configurations are summarized in the SM.
 \label{Fig:spin_pol}}
\end{figure}

The results for the TSP are summarized in Fig.~\ref{Fig:spin_pol} and in the SM~\footnotemark[1].
We show the individual spin-resolved transmissions, $T_{\uparrow / \downarrow}$, and the TSP, $P_T$ for the six stacking configurations. We additionally average over these stacking configurations to get an estimate for the total values to be expected in the medium and large supercell.

Looking at the results for S$_{1,2}$ which is the lowest-energy configuration, see Fig.~\ref{Fig:spin_pol}, we find that $T_{\uparrow}$ essentially follows the spin-up DOS of the Co lead \footnotemark[1]. Since the spin up DOS of hcp-Co is small and nearly constant for $E-E_F$ above -0.5 eV, also $T_{\uparrow}$ is small and constant for these energies. At lower energies, the large spin-up DOS of Co provides many incoming modes at many different $k$-vectors, see the movie on the spin- and $k$-resolved DOS of the Co lead (\texttt{spin\_and\_k\_resolved\_DOS\_hcp\_Co.mp4}), that can potentially tunnel through the Co/hBN/graphene/hBN/Co scattering region. From the transmission it is evident that these modes are supported by the scattering region. In contrast, $T_{\downarrow}$ is dominated by distinct peaks at certain energies. 

In the Supplemental Material~\footnotemark[1], we show movies (\texttt{trans\_movie\_config.mp4}) of the spin- and $k$-resolved transmissions as function of energy for the different stacking configurations. The total transmission for each spin-channel is given above the subfigures, which is a weighted sum of $k$-resolved transmissions within the Brillouin zone. The value of spin polarization, $P_T$, is also given for each energy. From there, it is evident that the peaks in $T_{\downarrow}$ are tunneling events near the Brillouin zone corners. In other words, the Brillouin zone spin filter mechanism for Co minority carriers produces resonance peaks. 

\begin{figure}[!htb]
	\includegraphics[width=0.99\columnwidth]{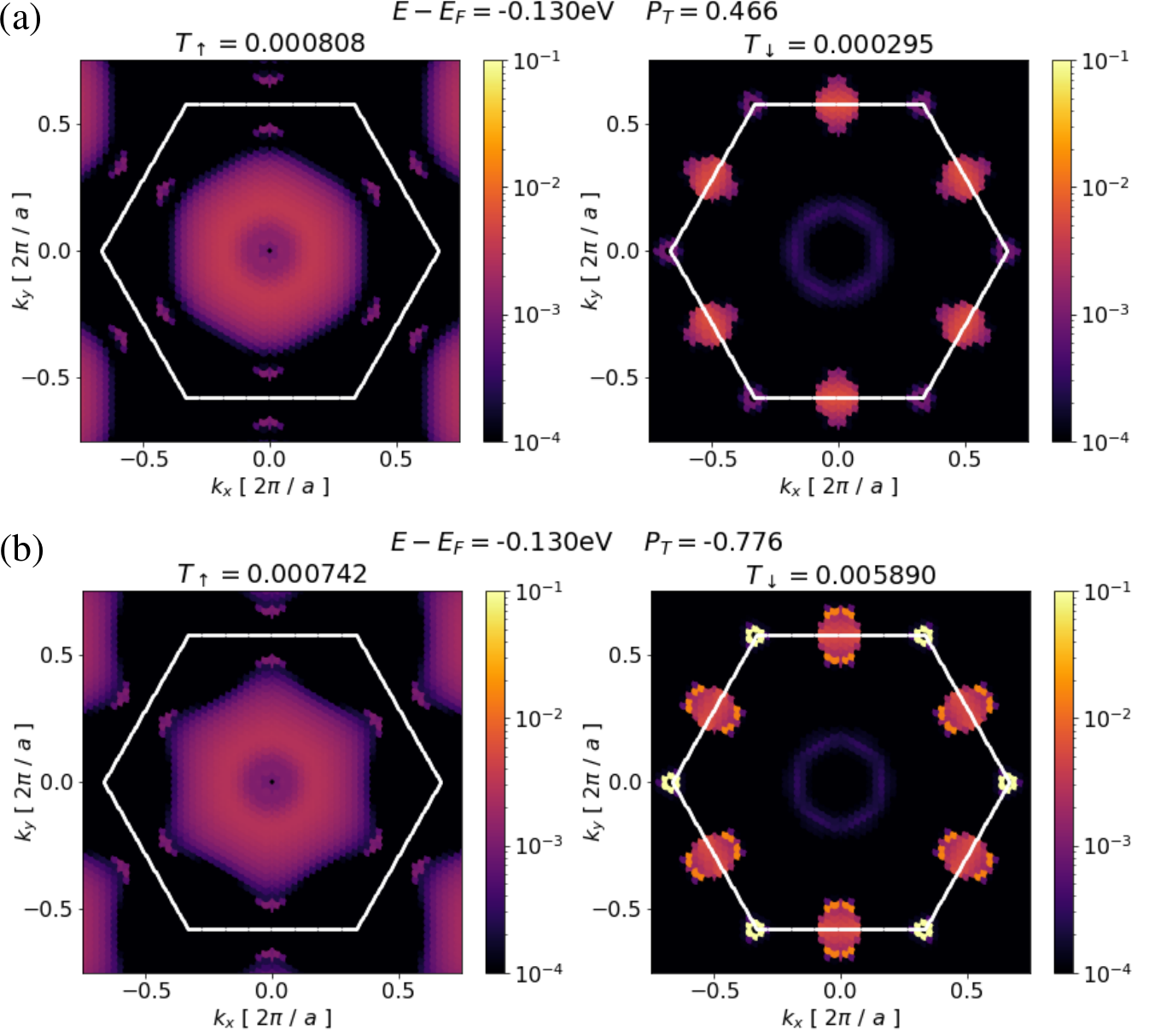}
	\caption{(a) Spin- and $k$-resolved transmissions for the S$_{1,2}$ scattering region for $E-E_F = -0.13$eV. Left (right) panel is for $T_{\uparrow}$ ($T_{\downarrow}$). The color code represents the magnitude of the transmission. The numbers for $P_T$ and $T_{\uparrow/\downarrow}$ above are total values at this particular energy. The white hexagon represents the edges of the Brillouin zone. (b) Same as (a), but for the S$_{2,2}$ scattering region.
 \label{Fig:compare_PT}}
\end{figure}

For clarification, in Fig.~\ref{Fig:compare_PT}, we compare the spin- and $k$-resolved transmissions, $T_{\uparrow / \downarrow}$, for S$_{1,2}$ and S$_{2,2}$ scattering regions for $E-E_F = -0.13$eV. 
We note that the overall $T_{\uparrow}$ for both scattering regions is rather similar, while there are pronounced differences in $T_{\downarrow}$, at this energy.
Compared to S$_{1,2}$, S$_{2,2}$ shows $T_{\downarrow}$ hot spots near the Brillouin zone corners, which are responsible for the strong resonance in $P_T$, see Fig.~\ref{Fig:spin_pol}.

Overall, while $T_{\uparrow}$ is rather similar for the different stacking configurations, $T_{\downarrow}$ sensitively depends on the atomic alignment, supporting the resonance modes at different energies. 
Even though the resonances are suppressed in the configuration average, their local nature provides evidence of spin tunneling \emph{hot spots}.

%------------------------------------------------------------------------
%------------------------------------------------------------------------
%------------------------------------------------------------------------
%------------------------------------------------------------------------

%------------------------------------------------------------
\section{Tailoring the atomic site overlap}
%------------------------------------------------------------
\label{Sec:Atomic_Site_Overlap}

The coupling between the layers comes primarily from vertically arranged atomic sites and the respective overlap of the $z$-extended orbitals ($p_z$ of C and N, $d_{z^2}$ of Co). To screen for maximized proximity coupling, mediated by the overlap between these orbitals, we consider the lattices of unstrained graphene, hBN, and a single layer of Co, and calculate the total and atom-resolved site overlap as a function of the two twist angles: one between graphene and hBN, the other between hBN and Co.

\begin{figure*}[htb]
    \centering
    \includegraphics[width=0.99\textwidth]{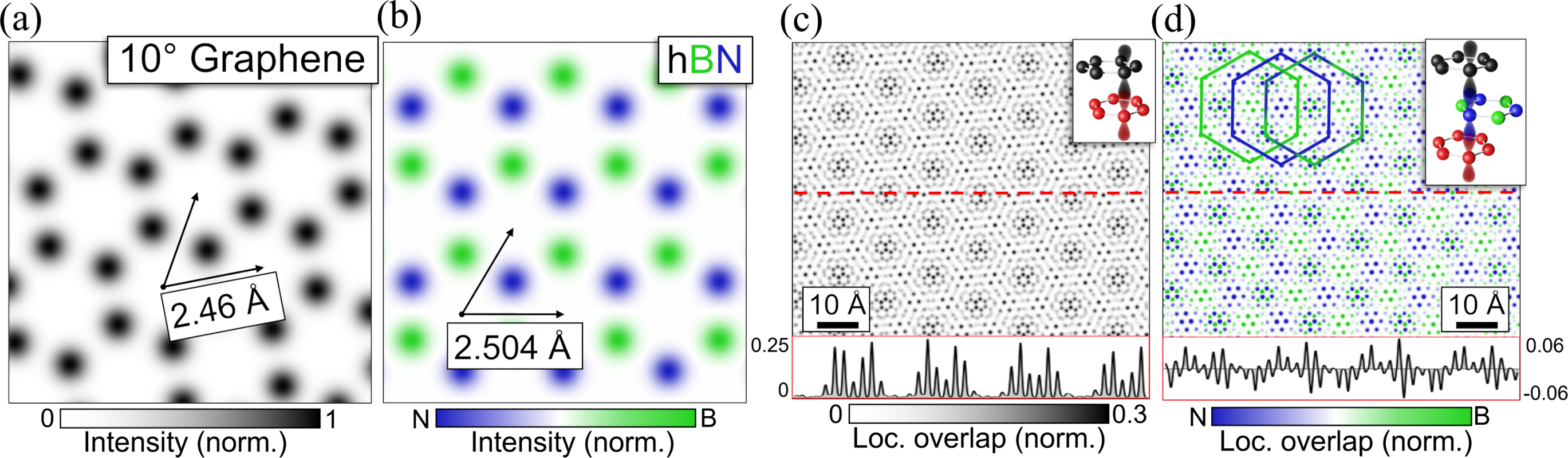} 
    \caption{(a) Top view of 10° rotated graphene, convolved with the Gaussian filter. (b) hBN lattice, where the B (N) atom is color coded green (blue), entering the overlap calculation as positive (negative) values. (c) Local overlap between 10° rotated graphene and 0° Co, showing moiré type patterns. Bottom insert shows a line-cut (red dashed line) of the overlap across the sample. (d) Local overlap between 10° graphene, 0° Co, and 0° hBN. Colored hexagons highlight 6-fold patterns for different atomic species (blue for N and green for B). The line-cut shows positive (negative) values indicating predominant vertical coupling through B (N) atoms.}
    \label{fig:overlap_expl}
\end{figure*}

We assess the overlap of the $z$-extended orbitals using a geometric approach. 
Each layer $\alpha$ (graphene, hBN, and Co) is simulated by a 2D array $L^{\alpha}_{ij}$, where 200$\times$200 pixels in a hexagonal arrangement are assigned a value of "1" to symbolize the lattice sites, corresponding to about 25$\times$25 nm of a real space lattice. The lattice sites are then blurred by a Gaussian ($\sigma=0.3  \textrm{\AA}$) to emulate the spatial distribution of the orbitals, see Fig.~\ref{fig:overlap_expl}(a,b). The matrices representing each material are then multiplied element-wise (Hadamard product) to obtain the local overlap. Say, for hBN and Co layers the overlap matrix would be $\Omega_{ij}=L^{\rm hBN}_{ij} \odot L^{\rm Co}_{ij}$. 
As an example, in Fig.\ref{fig:overlap_expl}(c) we show $\Omega_{ij}$ for a Co layer and a 10° twisted graphene layer, revealing 6-fold symmetric moiré physics. The calculated $\Omega$ is also shown for a line-cut across the sample, representative of the locally varying proximity coupling between the layers. 

%------------------------------------------------------------
%------------------------------------------------------------

In spin-injection geometries, we additionally need to consider the hBN barrier. As argued above, the proximity coupling between graphene and cobalt is then mediated primarily by the N $p_z$ orbitals.
To illustrate the effect of monolayer hBN as a \textit{filter} for proximity exchange, we weigh the contribution of the B (N) atom in the hBN lattice positive (negative) to distinguish the proximity coupling via the two atoms. The resulting $\Omega$ is shown in Fig.~\ref{fig:overlap_expl}(d).

Adding a monolayer hBN changes the local pattern significantly. While the 6-fold symmetry of Fig.~\ref{fig:overlap_expl}(c) is maintained in the filtered overlap of (d), in the trilayer stack the overlap occurs alternatively across different atomic species. We can identify regions where the proximity is mainly mediated by either N, B, or a mixture. Three respectively colored hexagons highlight these emerging sub-superlattices.
This simple approach can be extended to any combination of 2D-lattices comprised of $z$-extended orbitals, to get an idea of the proximity coupling.

\begin{figure}[htb]
    \centering
    \includegraphics[width=0.99\columnwidth]{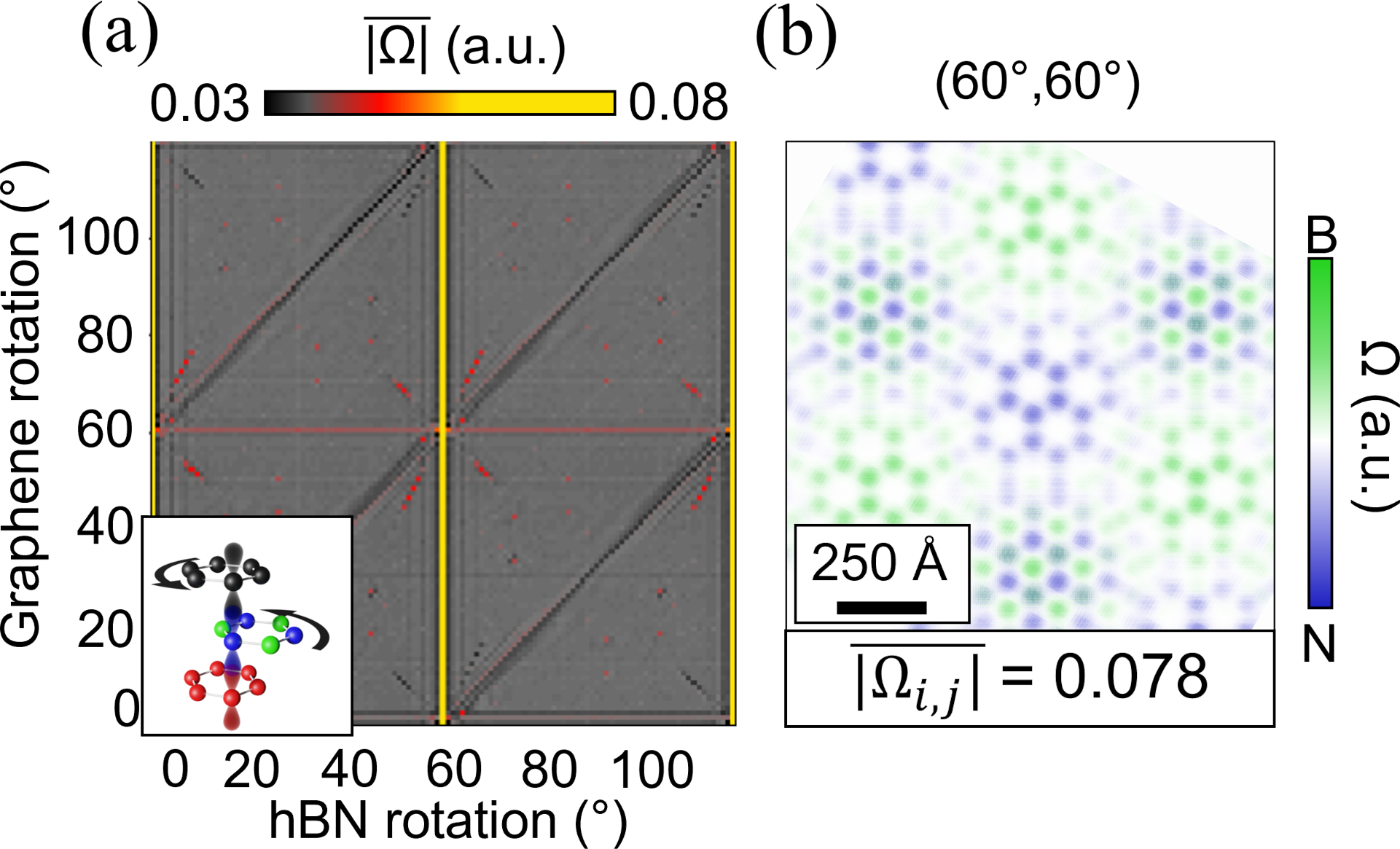} % 
    \caption{(a) Angle resolved mean absolute overlap $\overline{|\Omega_{i,j}|}$ in a 0° to 120° range for both graphene and hBN rotation. Plotted values are normalized to the value corresponding to three identical lattices with no twist. The inset shows each simulated material aligned with respect to one lattice site acting as a common rotation center. (b) $\Omega_{i,j}$ for the (60°,60°) twist angle combinations of graphene and hBN, corresponding to the highest mean absolute overlap value. Colors indicate vertical alignment with B (N) atoms.}
    \label{fig:overlap_res}
\end{figure}

Moreover, we can use the total mean absolute overlap $\overline{|\Omega_{ij}|}$, considering the whole simulated sample, as a measure of the proximity interaction.
Individually varying both twist angles in the trilayer stack, we can then identify regimes of maximal overlap. 
We simulate $\Omega$ of a graphene/hBN/Co stack --- omitting the B atoms of the hBN lattice as they do not mediate the proximity coupling --- while twisting the graphene and hBN layers from 0° to 120°. The inset of Fig.~\ref{fig:overlap_res}(a) shows how twist is defined: all simulated materials are constructed with a single coincident lattice site, which acts as the center of rotation. Note that the choice of the center of rotation does not affect our results, as any other center of rotation would correspond to the same rotation combined with a lateral translation. It can be shown that the $\overline{|\Omega_{ij}|}$ of unequal lattices is translation invariant. 

The results of the simulation are shown in Fig.~\ref{fig:overlap_res}(a). This plot shows the angles at which the total overlap between C and Co (via N) is maximized. We notice that for all twist angles of the hBN layer that are multiples of 60°, we obtain high overlap values. This is to be expected as the lattice constant of hBN and hcp-Co are very similar. High overlap values are also achieved for graphene twist angles that are multiples of 60°. However, due to the greater lattice disparity between graphene and Co, the overlap is smaller.
Along the diagonal, we find the angles where the twist of graphene equals that of hBN. Interestingly, we find that the overlap is reduced, likely due to the lattice mismatch between graphene and hBN, combined with the twist relative to the Co layer. Finally, a few unexpected high-overlap locations appear around the points where the horizontal and vertical strong-overlap lines intersect. 
Fig.~\ref{fig:overlap_res} (b) displays the local overlap for a representative twist angle configuration. Like in Fig. \ref{fig:overlap_expl}, the color indicates the overlap mediated by either N or B. Albeit on a much larger scale, the 6-fold patterns described in Fig.~\ref{fig:overlap_expl}(d) appear again.

\emph{From these observations follows that the best geometry for maximal exchange proximity effect is obtained by keeping hBN aligned with Co. The alignment of graphene is then not critical.} The effective overlap for aligned hBN/Co is three times stronger than for not aligned structures, as seen in Fig.  \ref{fig:overlap_res}.
Moreover, we identified other regions that, albeit showing a reduced overlap, present non-trivial superlattice structure, both in periodicity and nature of the coupling mediation through the hBN layer, which can be leveraged for studying the effect of moiré structures (see Supplementary F).

%------------------------------------------------------------
\section{Summary and Conclusion}
%------------------------------------------------------------
\label{Sec:Summary}
We presented a systematic theoretical investigation of local proximity effects in Co/hBN/graphene heterostructures, focusing on their role in spin injection for 2D spintronic devices. The main findings are (i) Locality of Proximity Exchange: The induced exchange splitting in graphene's Dirac bands varies from 1 to 100~meV depending on the local atomic stacking. Proximity effects are highly localized and sensitive to geometry; (ii) Orbital Hybridization Mechanism: Proximity exchange is driven by hybridization between Co $d_{z^2}$, hBN $p_z$, and graphene $p_z$ orbitals. The degree of wavefunction compatibility governs the strength of the effect;
(iii) Variation in Spin Transport Properties: Different stackings lead to diverse induced magnetic moments and tunneling spin polarizations (TSPs), revealing spatial inhomogeneities and local spin-filtering behavior;
(iv) Effect of Additional Layers and Twist: Inserting extra hBN or graphene layers and introducing twist angles modulates proximity exchange and TSP, offering further control over local spin properties;
(v) Implications for Device Design: Our results emphasize that precise control of stacking and interlayer alignment is crucial for optimizing spin injection and interpreting experimental measurements in vdW heterostructures.

\acknowledgments
K.~Z., L.~C., and J.~F. were supported by the Deutsche Forschungsgemeinschaft (DFG, German Research Foundation) SFB 1277 (Project No. 314695032), SPP 2244 (Project No. 443416183), the EU  project 2DSPIN-TECH (Project No. 101135853), and FLAGERA project 2DSOTECH. R.~S. and A.~V.~S. acknowledge the DFG for financial support via the Priority Programmes SPP 2244, as well as the clusters of excellence MCQST (EXS-2111) and e-conversion (EXS-2089). 
    
\footnotetext[1]{See Supplemental Material, including Refs. \cite{Constantinescu2013:PRL, Mazin1999:PRL, Patel2023:PRB} where we provide additional results on DFT-calculated band structures and density of states, overlap simulations, and exemplary input files. }

\bibliography{references}

\cleardoublepage


\section*{Supplementary material}

# Supplemental Material: Resonant magnetic proximity hot spots in Co/hBN/graphene

Klaus Zollner[1],[*] Lukas Cvitkovich[1], Riccardo Silvioli[2], Andreas V. Stier[2], and Jaroslav Fabian[1],[3]

[1]*Institute for Theoretical Physics, University of Regensburg, 93053 Regensburg, Germany*

[2]*Walter Schottky Institut and TUM School of Natural Sciences,*

*Technische Universität München, 85748 Garching, Germany and*

[3]*Halle-Berlin-Regensburg Cluster of Excellence CCE, Germany*

## I. RESULTS

In the following, we present additional calculation results, supporting the main text. In particular, we present DFT-calculated dispersions and density of states for all considered stacking configurations, exemplary dispersions with spin-orbit coupling, results for additional hBN and graphene layers, comparison of spin polarization definitions, and exemplary input files.

* klaus.zollner@physik.uni-regensburg.de

## A. Monolayer Graphene, Monolayer hBN

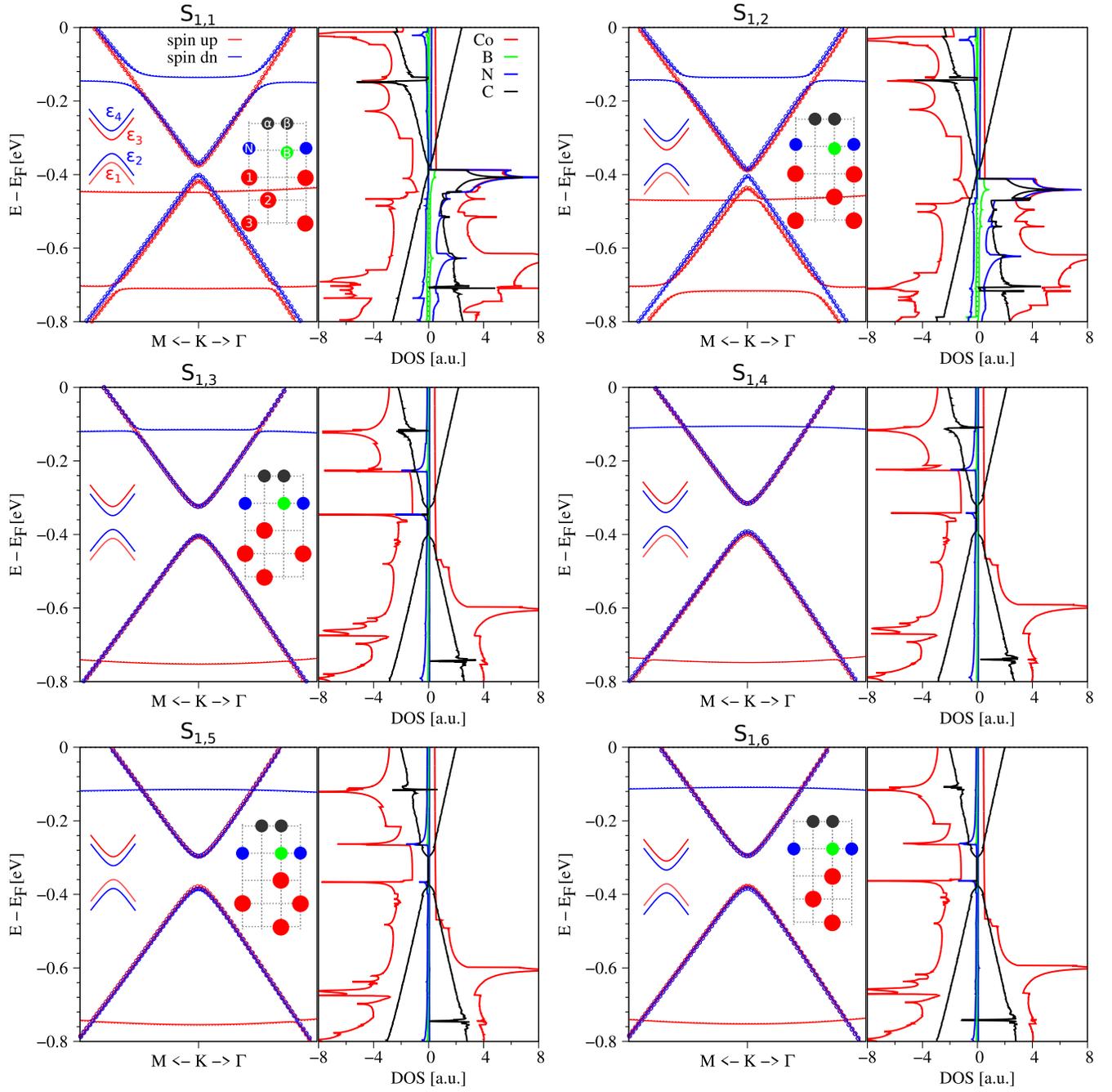

FIG. 1. DFT-calculated dispersions, and spin- and atom-resolved density of states, in the vicinity of the Dirac point for the stacking configurations $S_{1,1}$ - $S_{1,6}$. For the band structure, red and blue correspond to spin up and spin down. In the insets we show the stacking geometries, respectively. The red and blue open circles represent the projection on graphene orbitals. In the insets, we also sketch the low energy Dirac bands which are split into four eigenvalues $\varepsilon_1 - \varepsilon_4$. For better visualization, the B/N DOS is multiplied by a factor of 10, while the C DOS is multiplied by a factor of 100.

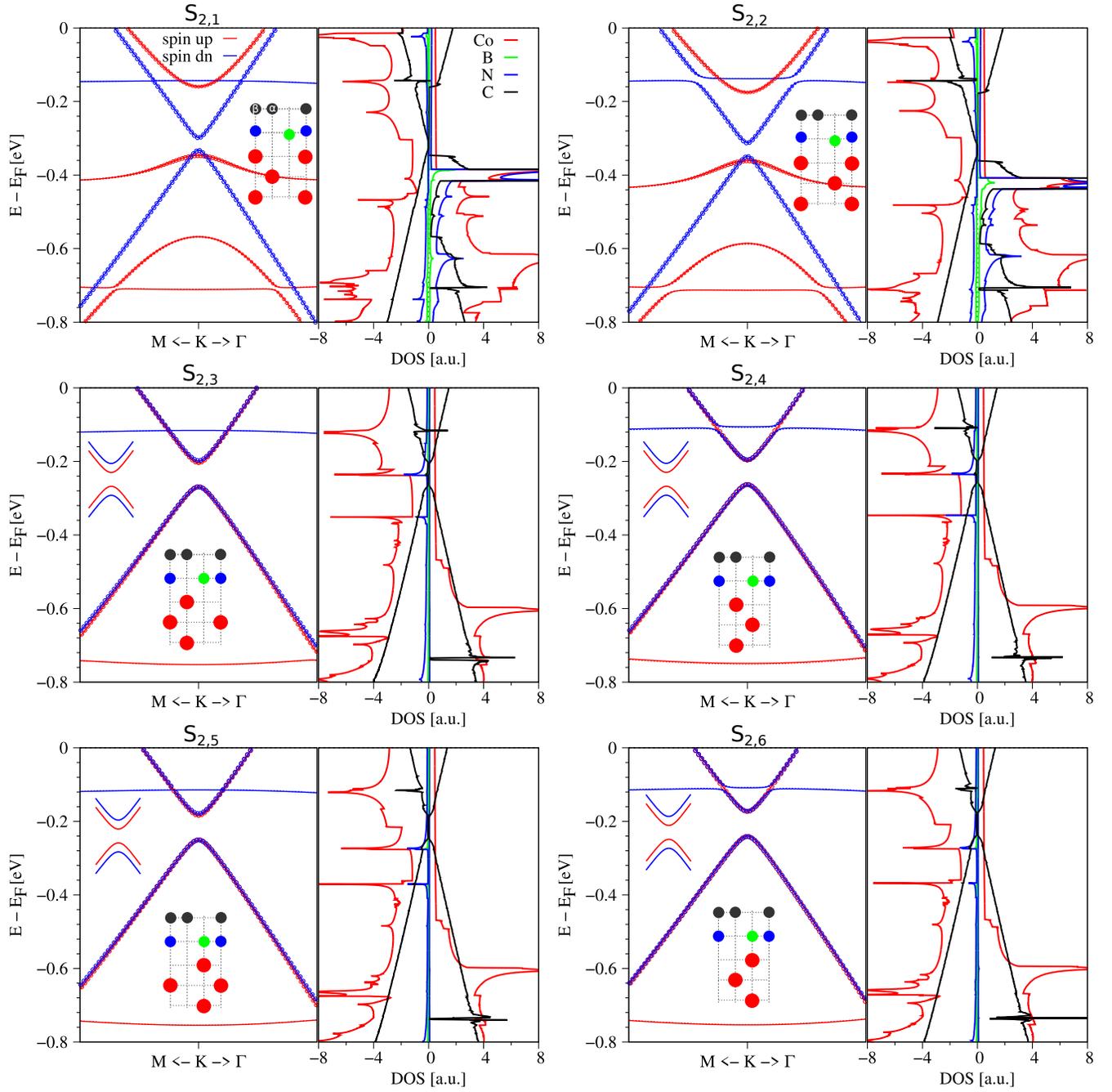

FIG. 2. Same as Fig. 1, but for the stacking configurations $S_{2,1}$ - $S_{2,6}$.

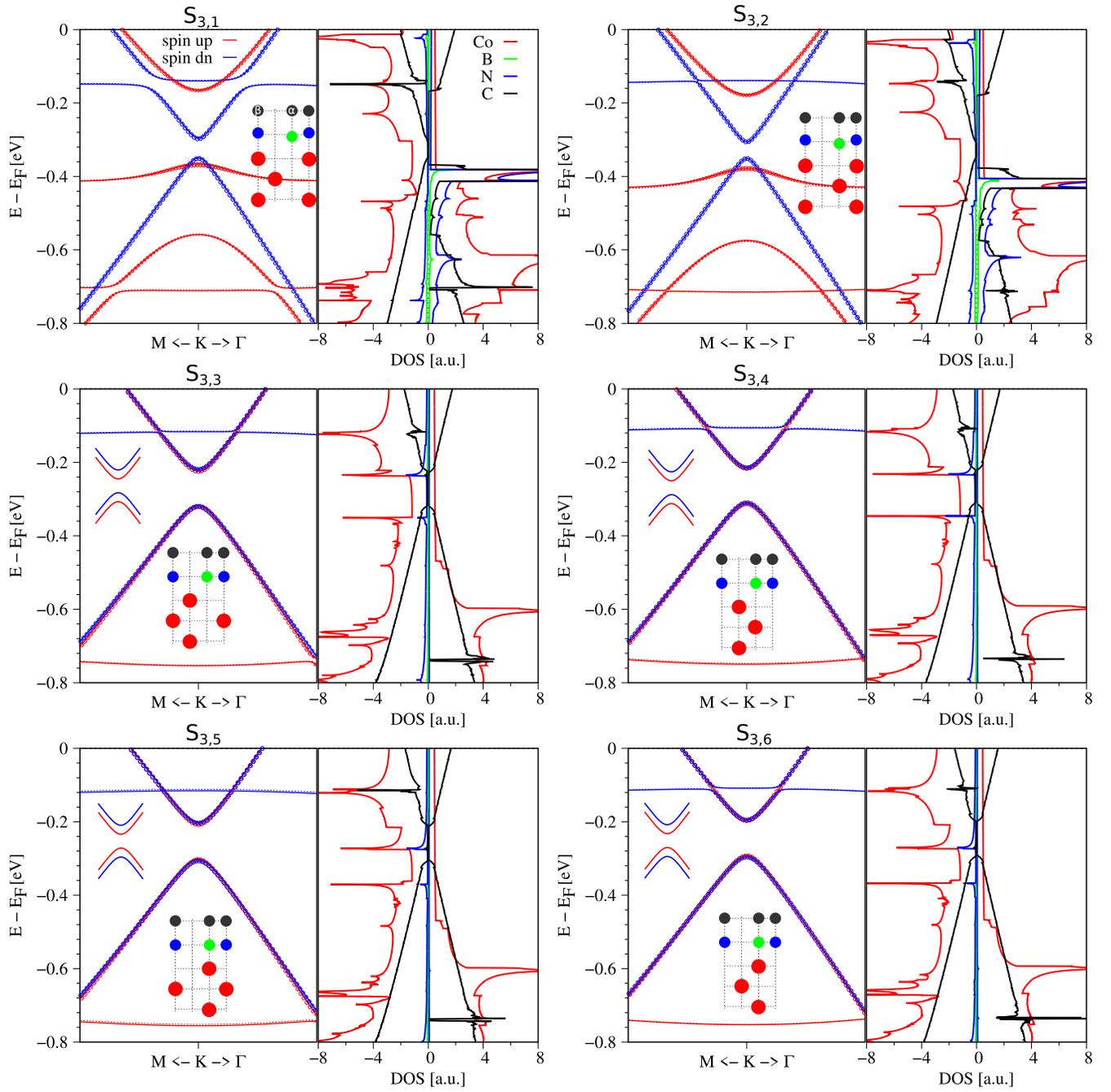

FIG. 3. Same as Fig. 1, but for the stacking configurations $S_{3,1}$ - $S_{3,6}$.

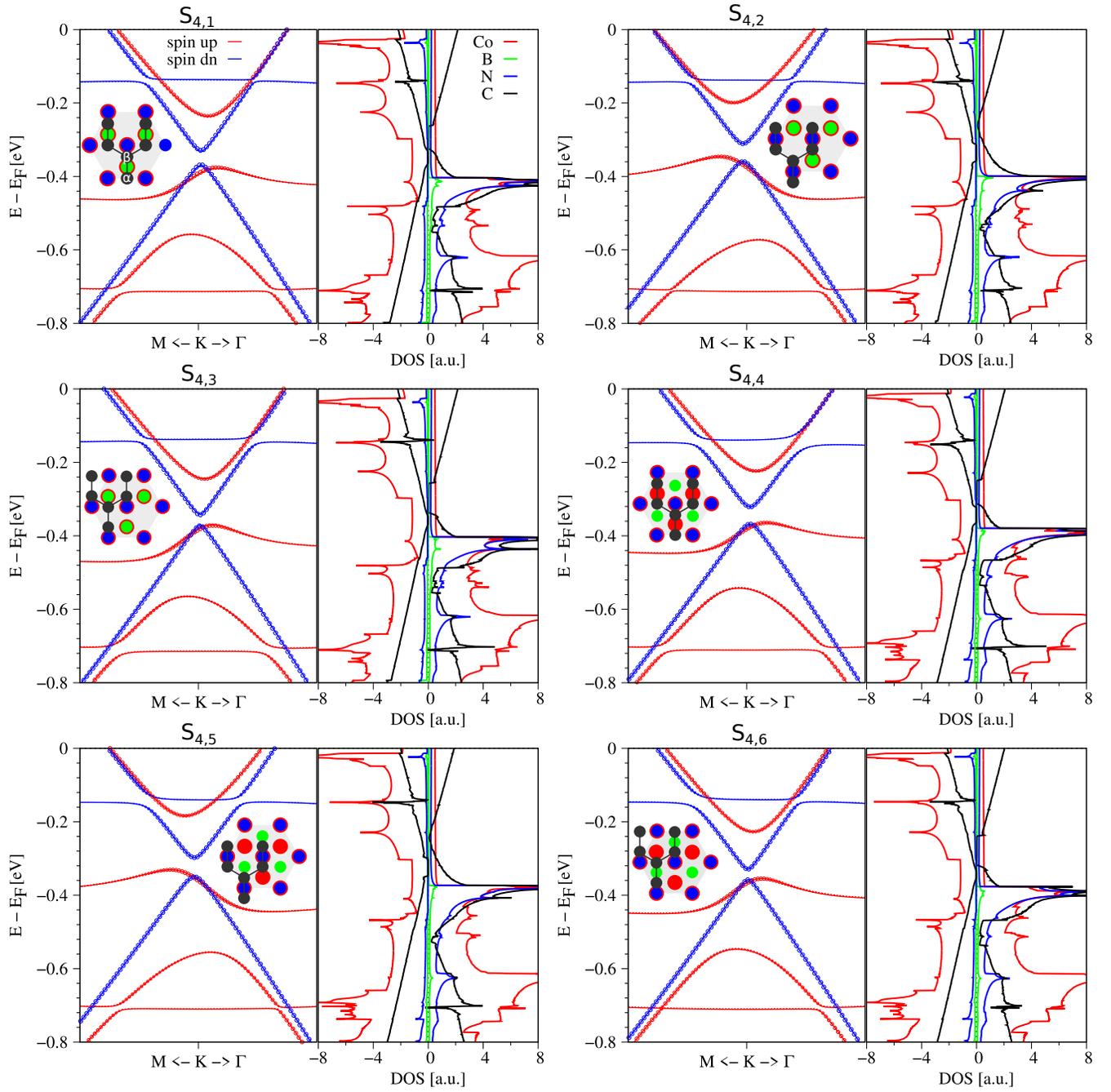

FIG. 4. Same as Fig. 1, but for the stacking configurations $S_{4,1}$ - $S_{4,6}$.

$S_{1,2}$            $S_{2,2}$

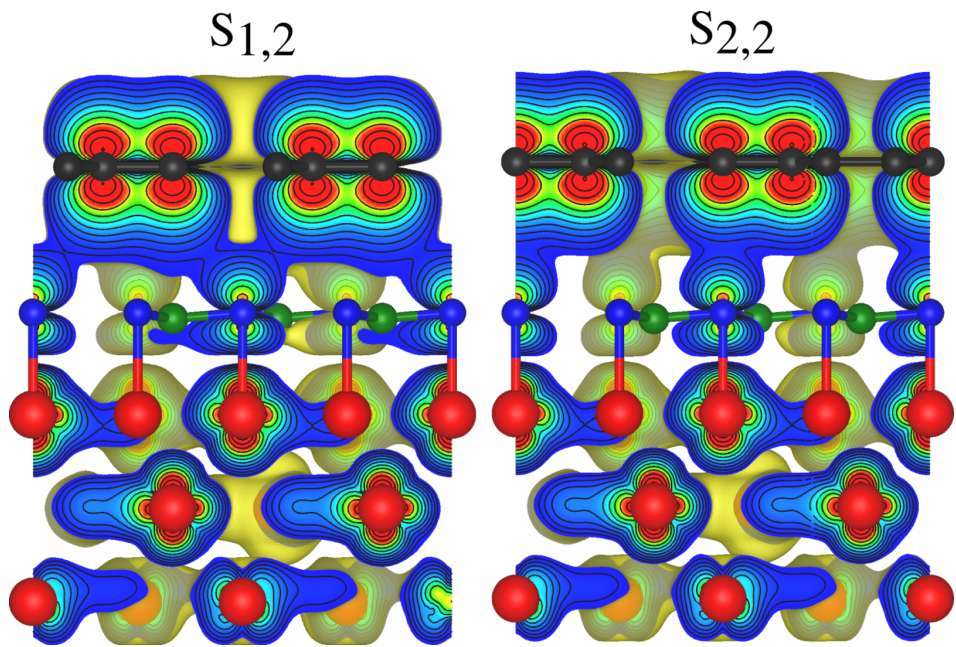

FIG. 5. DFT-calculated integrated local density of states, emphasizing the hybridization channels from Co $d_z^2$ to C $p_z$ orbitals via N $p_z$ orbitals for two exemplary stackings.

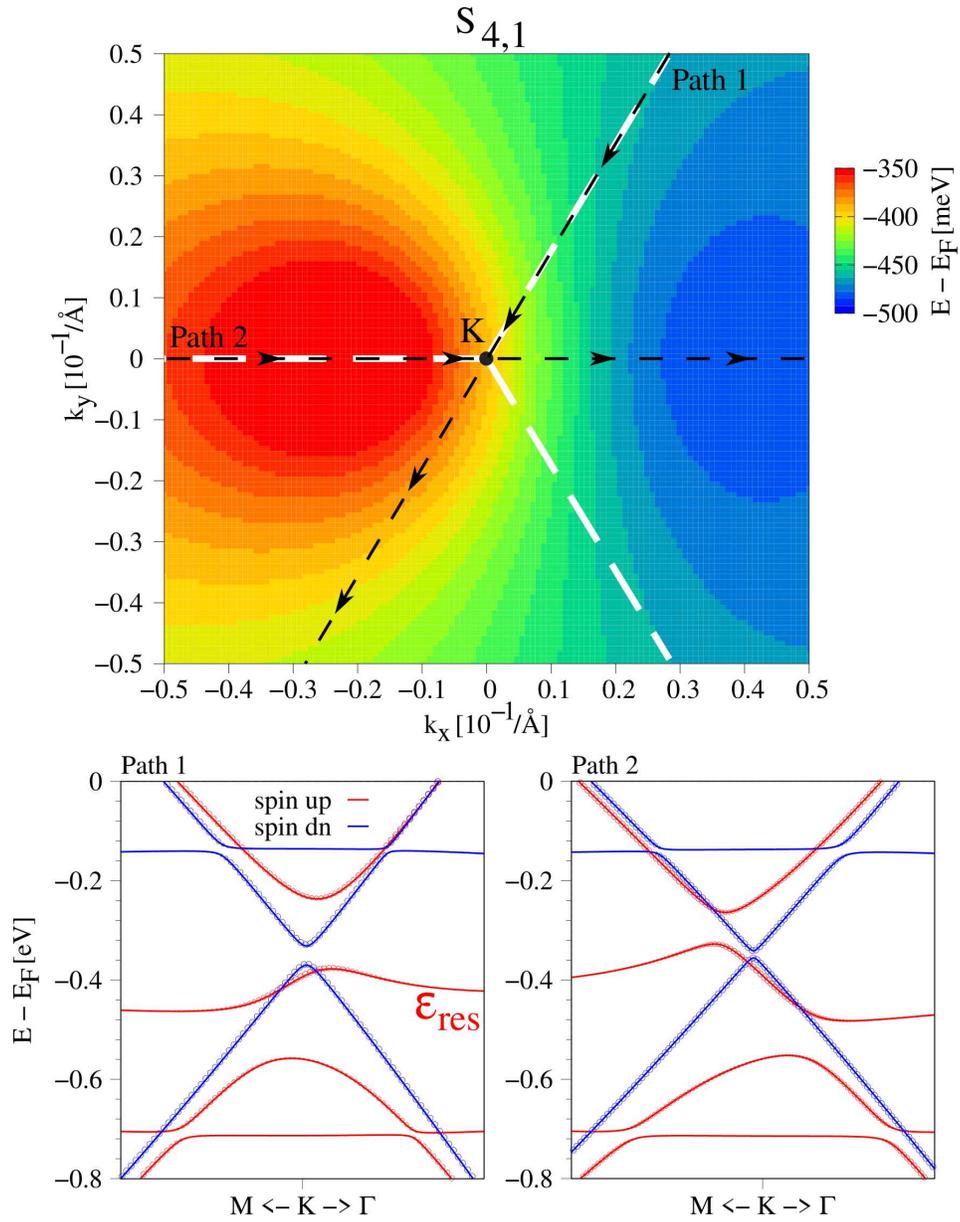

FIG. 6. 2D-map of the resonant energy level $\varepsilon_{res}$ in the vicinity of the K point for the stacking $S_{4,1}$. The color code indicates the energy with respect to the Fermi level. The white dashed lines indicate the edge of the Brillouin zone. Below, we show low energy dispersions along 2 different paths, as indicated in the 2D-map. We also indicate the resonant energy level in the dispersion. This demonstrates the anisotropy of the band hybridization for low-symmetry stackings.

In the calculations of the main text, we have omitted the effects of spin-orbit coupling. To justify this, we consider two structures, $S_{1,2}$ and $S_{2,2}$, and two different spin quantization directions, along $\langle 001 \rangle$ (z-direction, perpendicular to the layers) and $\langle 100 \rangle$ (x-direction, parallel to the layers). For the calculations, we consider the relativistic versions of the pseudopotentials and employ a $k$-grid of $120 \times 120 \times 1$. In Figs. 7 and 8 we compare the proximitized low energy Dirac bands for the different stackings and quantization directions. From the total energies of the $S_{1,2}$ ($S_{2,2}$) stackings, we find that the $\langle 001 \rangle$ direction is energetically favorable by about 40 $\mu$eV for both configurations. This behavior may be the due to the very thin Co, where surface anisotropy dominates the magnetization, leading to an out-of-plane easy axis [1]. Experimentally, one would rather expect an in-plane easy axis for thicker hcp-Co slabs. Only for the $\langle 001 \rangle$ magnetization, we see a notable difference in the proximitized low energy Dirac bands near K and K', indicting the interplay of proximity exchange and spin-orbit coupling. Apart from that, spin-orbit coupling does not play a significant role for the proximity effect.

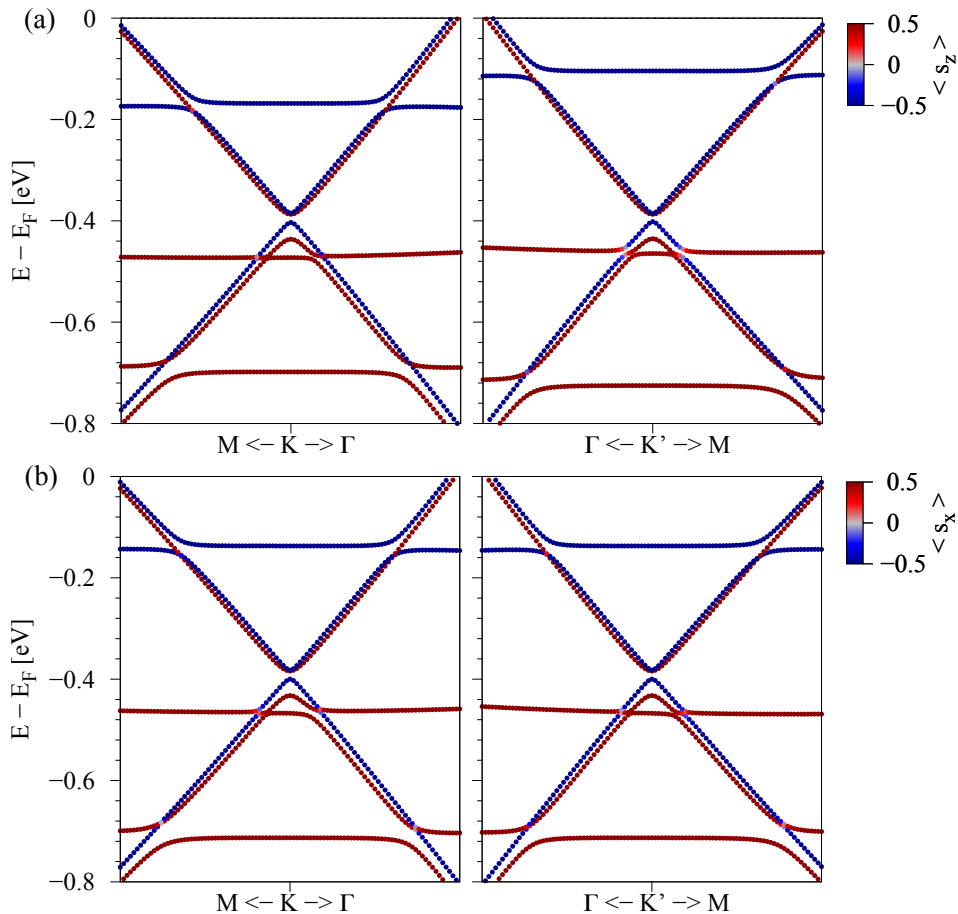

FIG. 7. DFT-calculated dispersions in the vicinity of the Dirac point near K (left) and K' (right) for the $S_{1,2}$ stacking configuration. Here, spin-orbit coupling is included. (a) The magnetization is along $\langle 001 \rangle$ direction, while in (b) it is along $\langle 100 \rangle$ direction. The color codes in (a) and (b) represent the projection of the spin parallel to the magnetization, respectively. Even though, the spin-orbit coupling creates a slight valley-dependence, as Co $d$-levels show a spin-orbit-induced valley-dependence, the overall Dirac dispersion is mostly governed by proximity exchange via band anticrossings.

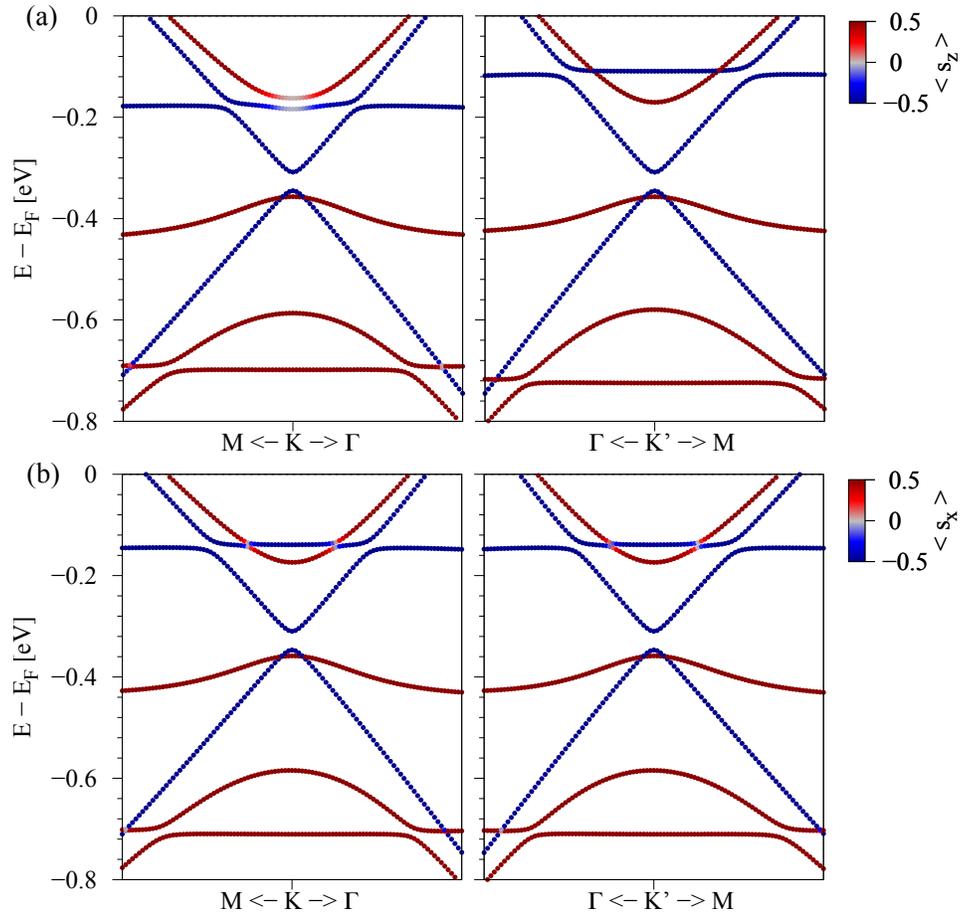

FIG. 8. Same as Fig. 7, but for the $S_{2,2}$ stacking configuration.

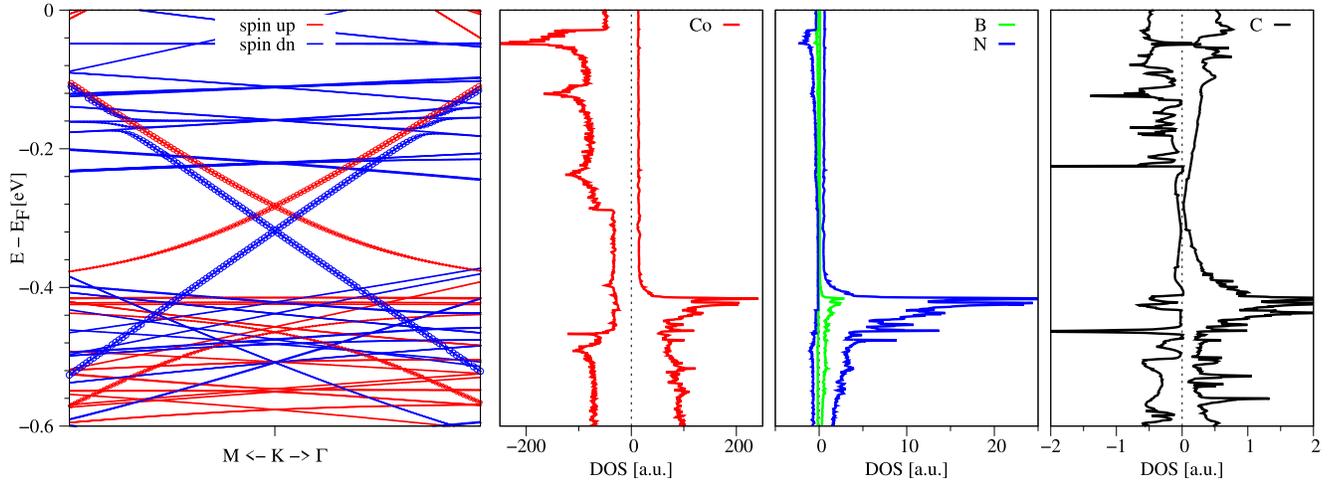

FIG. 9. DFT-calculated band structure in the vicinity of the K point and atom- and spin-resolved density of states for the medium-sized Co/hBN/graphene heterostructure. The geometry contains 191 atoms and is shown in the main text.

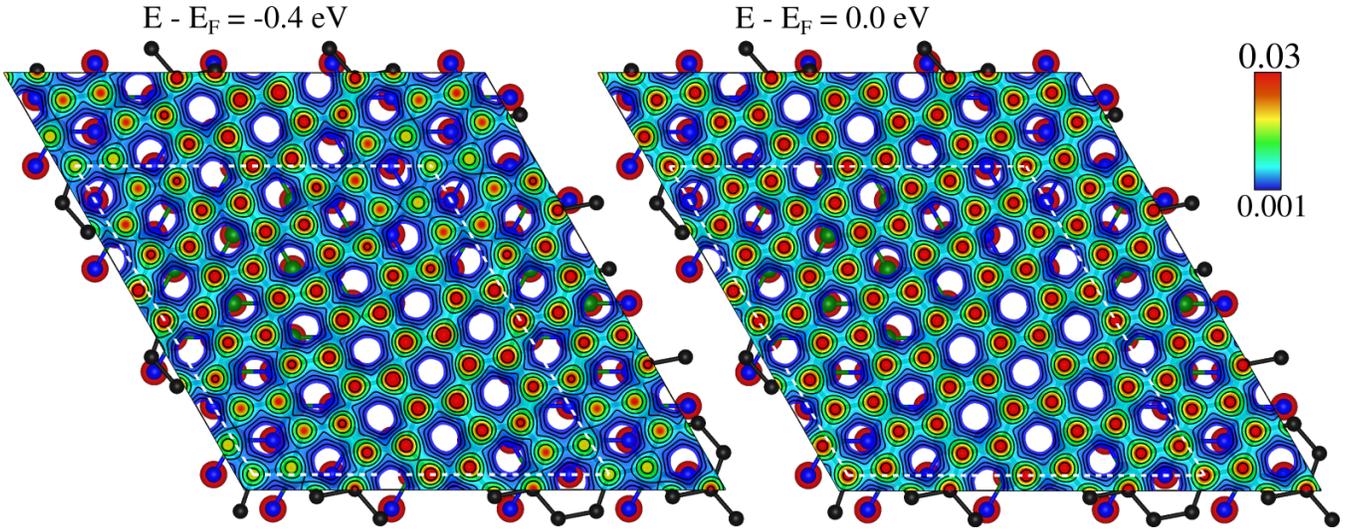

FIG. 10. DFT-calculated local density of states (LDOS) for the medium-sized Co/hBN/graphene heterostructure at different energies. We cut through the $p_z$-orbital density slightly above the C atoms. The white dashed line is the unit cell. The integration employs states at the given energy with 1 mRy of Gaussian broadening. The color scale corresponds to isovalues in units $e/\text{Å}^3$. At $-0.4$eV, the LDOS is non-uniform with suppressed contribution at the $S_{1,2}$ configuration site.

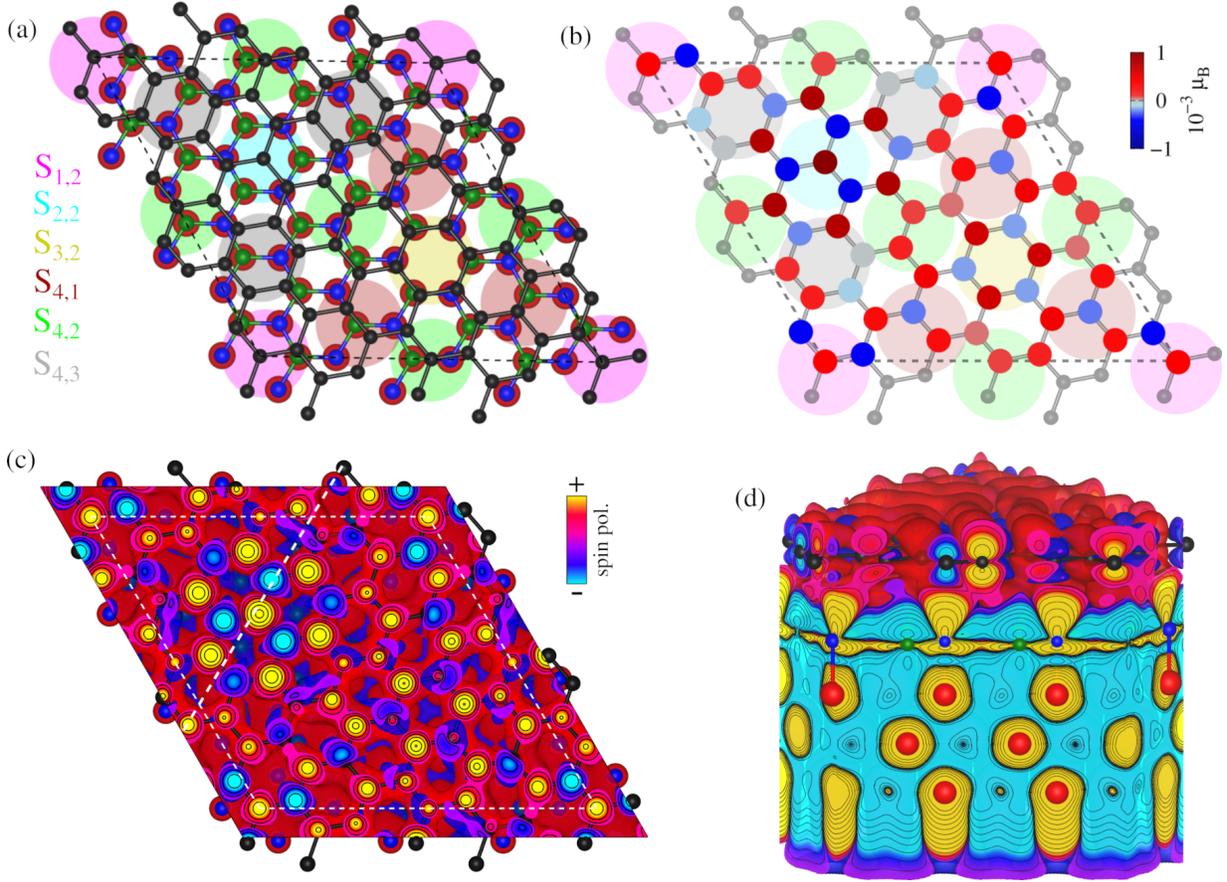

FIG. 11. (a) Medium-sized Co/hBN/graphene supercell. The different colored spheres emphasize the different local high-symmetry stackings. The dashed line represents the unit cell. (b) The proximity-induced calculated magnetic moments on C atoms, overlayed on the geometry. (c) DFT-calculated spin polarization for the medium-sized Co/hBN/graphene heterostructure. We cut through the $p_z$-orbital polarization slightly above the C atoms. Yellow/Red corresponds to positive polarization, while Cyan/Blue correspond to negative polarization, in line with the magnetic moments shown in (b). (d) Side view of (c), where we take a cut through the supercell, as indicated in (c).

We also briefly look into a larger geometry, see Fig.12. This is an aligned energetically most favorable Co/hBN interface, with a graphene layer on top at a twist angle of roughly 3.62°. This structure has lattice-matched Co/hBN, from the $S_{1,2}$ geometry, but with a lattice constant of 2.504 Å, resulting in a strain for Co of about -0.12%. The graphene layer in this supercell has a lattice constant of 2.4614 Å and is strained by about 0.057 %. The supercell comprises 1619 atoms. Although this system is too large for our methods to calculate the full electronic structure, we present it to illustrate the spatial variation of magnetic proximity effects arising from different local registries $S$. The colored circles indicate local stacking configurations resembling the lattice-matched registries $S$ from the main text.

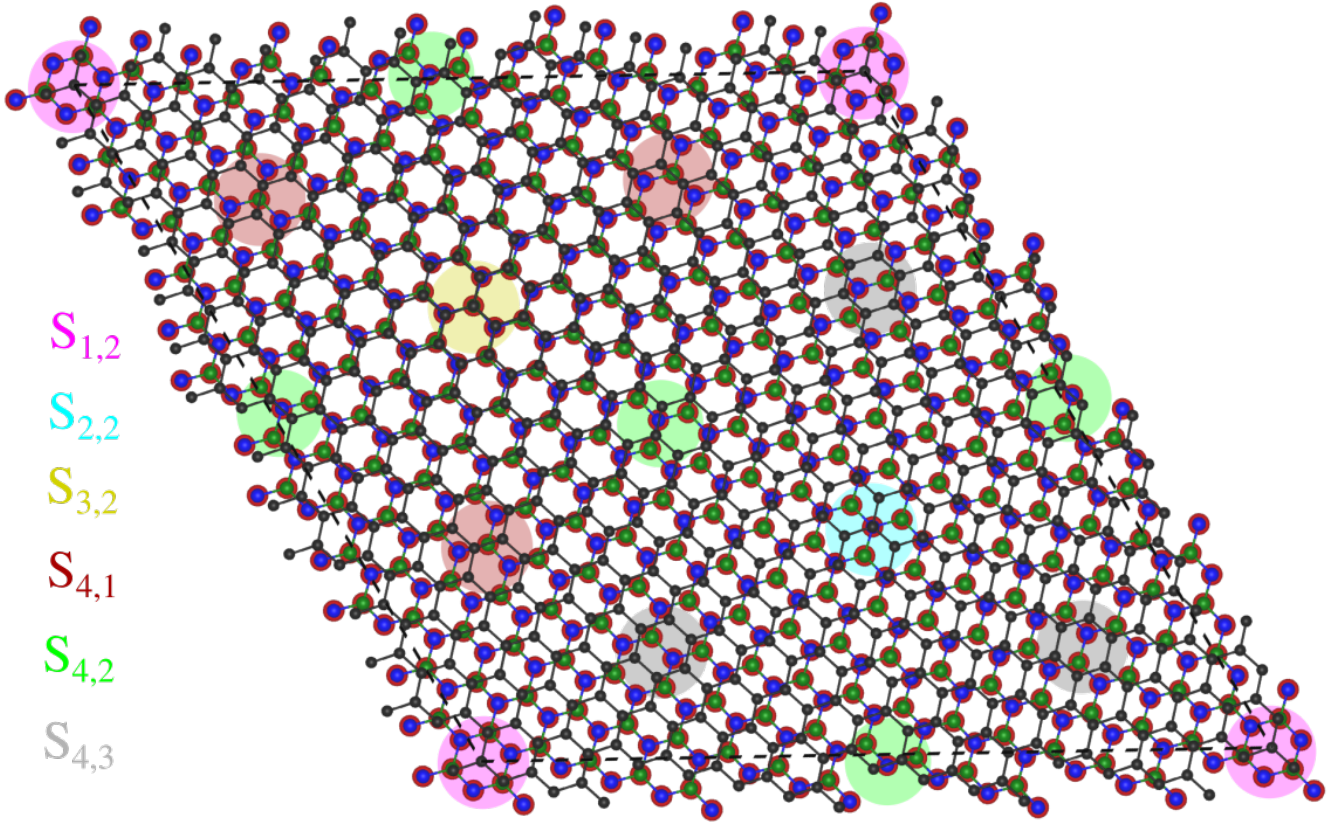

FIG. 12. Large unstrained Co/hBN/graphene supercell heterostructure. The different colored spheres emphasize the different local high-symmetry stackings. Unfortunately the cell is computationally too demanding.

## B. Monolayer Graphene, Bilayer hBN

We consider the six energetically most likely Co/hBN/graphene structures $S_{1,1}$, $S_{1,2}$, $S_{2,1}$, $S_{2,2}$, $S_{3,1}$, and $S_{3,2}$, and add another hBN layer between the existing hBN and graphene layers. The hBN stacking is considered to be AA' [2]. The results are summarized in Fig. 13 and Table I.

TABLE I. Total energies with respect to the lowest energy reference structure $S_{2,2}$, averaged interlayer distances between the layers, $d_{Co/hBN}$, $d_{hBN/hBN}$, and $d_{hBN/Gr}$, rippling of the lower hBN layer, $\delta_z$, and induced atomic magnetic moments

| config. | $E_{tot} - E_0$ [meV] | $d_{Co/hBN}$ [Å] | $\delta_z$ [Å] | $d_{hBN/hBN}$ [Å] | $d_{hBN/Gr}$ [Å] | $B_1$ [$10^{-3}\mu_B$] | $N_1$ [$10^{-3}\mu_B$] | $N_2$ [$10^{-3}\mu_B$] | $B_2$ [$10^{-3}\mu_B$] | $\alpha$ [$10^{-3}\mu_B$] | $\beta$ [$10^{-3}\mu_B$] |
|---|---|---|---|---|---|---|---|---|---|---|---|
| $S_{1,1}$ | 33.212 | 2.038 | 0.060 | 3.041 | 3.296 | -43.82 | 15.68 | 1.15 | -0.10 | -0.02 | 0.02 |
| $S_{1,2}$ | 29.365 | 2.036 | 0.060 | 3.042 | 3.295 | -47.19 | 13.47 | 1.10 | -0.23 | -0.03 | 0.03 |
| $S_{2,1}$ | 3.929 | 2.038 | 0.060 | 3.035 | 3.112 | -43.57 | 15.74 | 1.10 | -0.11 | -0.04 | 0.02 |
| $S_{2,2}$ | 0 | 2.036 | 0.060 | 3.036 | 3.110 | -46.92 | 13.55 | 1.04 | -0.22 | -0.05 | 0.03 |
| $S_{3,1}$ | 37.997 | 2.039 | 0.060 | 3.037 | 3.354 | -43.71 | 15.67 | 1.10 | -0.04 | 0.03 | 0.01 |
| $S_{3,2}$ | 34.128 | 2.036 | 0.060 | 3.039 | 3.352 | -47.08 | 13.47 | 1.04 | -0.16 | 0.03 | 0 |

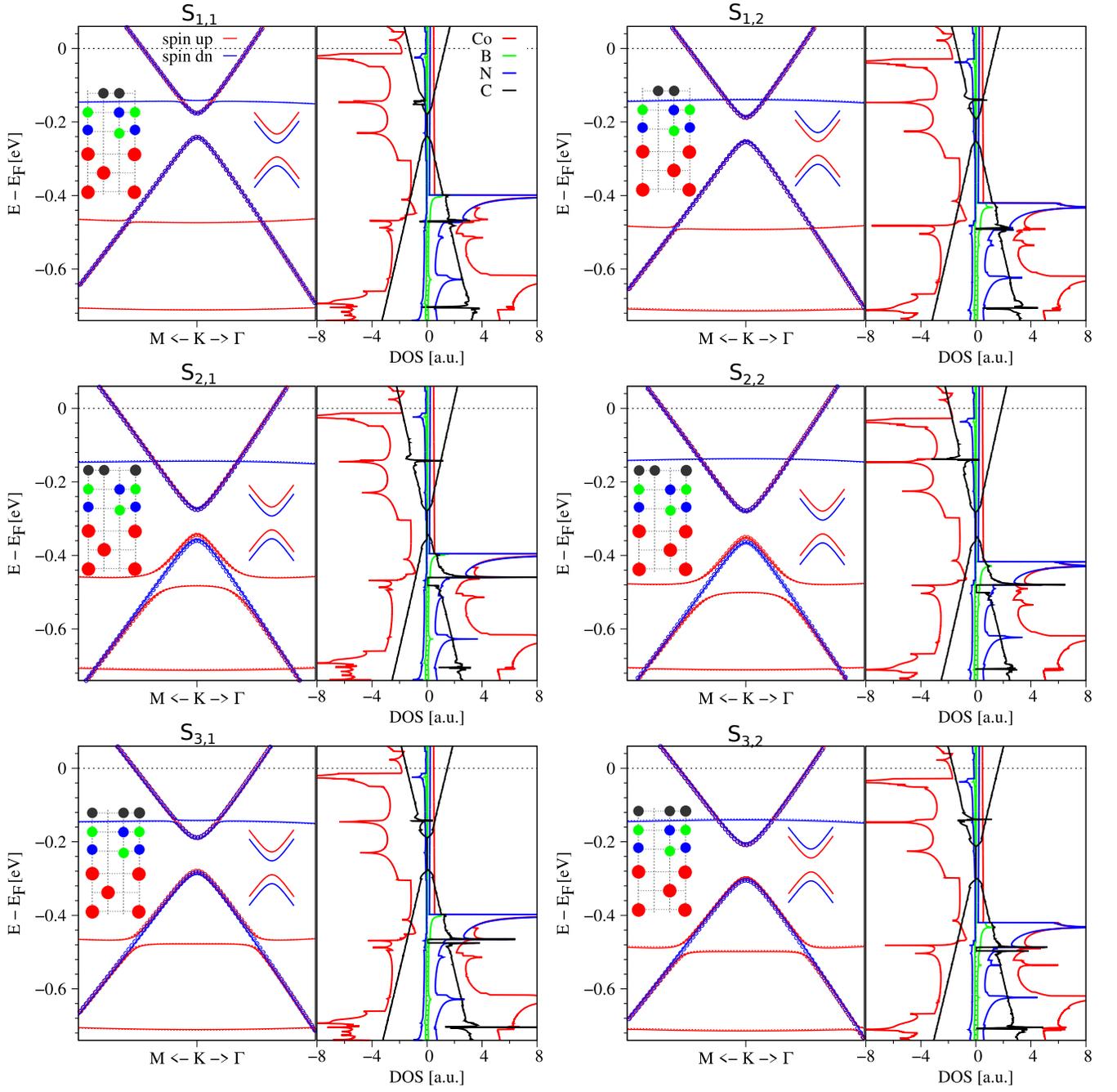

FIG. 13. Same as Fig. 1, but for the different graphene/bilayer-hBN/Co stacks.

### C. Bilayer Graphene, Monolayer hBN

We consider the three energetically most likely structures S$_{1,1}$, S$_{1,2}$, and S$_{2,2}$, and add another graphene layer on top. For each of the structures, there are two possibilities to add the graphene layer, such that we end up with Bernal bilayer graphene on the hBN/Co substrates. The results are summarized in Fig. 14 and Table II.

TABLE II. Total energies with respect to the lowest energy reference structure S$_{1,2}^1$, averaged interlayer distances between the layers, d$_{Co/hBN}$, d$_{hBN/Gr}$, and d$_{Gr/Gr}$, rippling of the lower hBN layer, $\delta_z$, and induced atomic magnetic moments

| config. | $E_{tot} - E_0$ [meV] | d$_{Co/hBN}$ [Å] | $\delta_z$ [Å] | d$_{hBN/Gr}$ [Å] | d$_{Gr/Gr}$ [Å] | B [$10^{-3}\mu_B$] | N [$10^{-3}\mu_B$] | $\alpha_1$ [$10^{-3}\mu_B$] | $\beta_1$ [$10^{-3}\mu_B$] | $\beta_2$ [$10^{-3}\mu_B$] | $\alpha_2$ [$10^{-3}\mu_B$] |
|---|---|---|---|---|---|---|---|---|---|---|---|
| S$_{1,1}^1$ | 4.323 | 2.042 | 0.058 | 3.060 | 3.242 | -41.82 | 15.37 | 0.51 | 0.13 | 0 | 0.03 |
| S$_{1,1}^2$ | 4.668 | 2.042 | 0.058 | 3.059 | 3.244 | -41.85 | 15.33 | 0.28 | 0.35 | 0.05 | -0.05 |
| S$_{1,2}^1$ | 0 | 2.040 | 0.058 | 3.060 | 3.241 | -44.96 | 13.15 | 0.40 | 0.18 | -0.03 | 0.04 |
| S$_{1,2}^2$ | 0.330 | 2.039 | 0.058 | 3.060 | 3.243 | -45.01 | 13.11 | 0.17 | 0.48 | 0.03 | -0.05 |
| S$_{2,2}^1$ | 27.076 | 2.035 | 0.060 | 3.188 | 3.241 | -45.39 | 12.47 | 3.76 | -2.74 | -0.45 | 0.46 |
| S$_{2,2}^2$ | 28.048 | 2.035 | 0.060 | 3.195 | 3.248 | -45.42 | 12.53 | 3.83 | -3.26 | -0.23 | 0.27 |

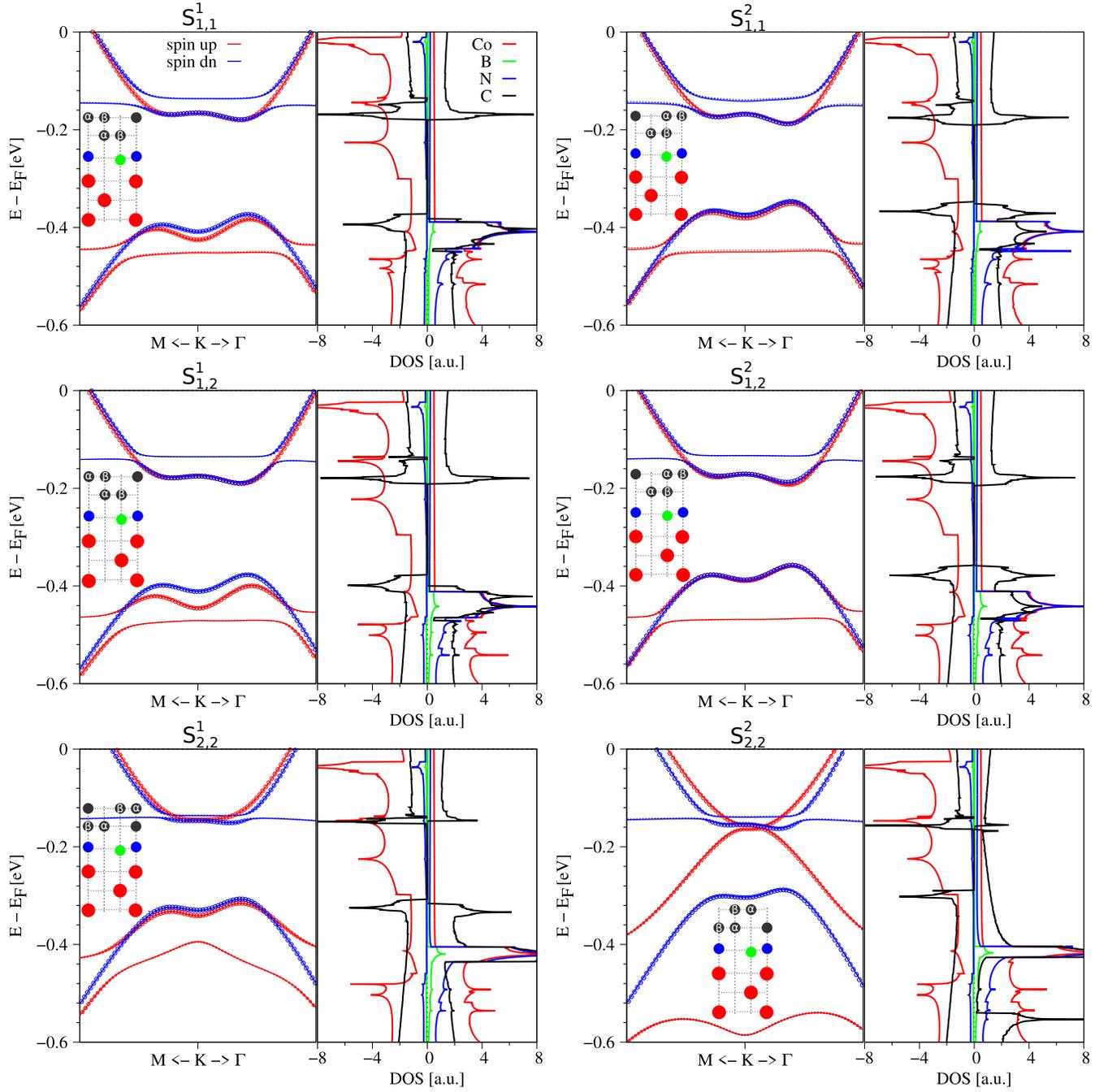

FIG. 14. Same as Fig. 1, but for the different bilayer-graphene/hBN/Co stacks. The C DOS is multiplied by a factor of 50.

### D. Definitions of Tunneling Spin Polarization

To get an estimate for the tunneling spin polarization (TSP), typically only the spin-resolved density of states (DOS) is considered. However, when we deal with heterostructures and interfaces, this definition may be misleading and lead to wrong interpretations of results, since Bloch band velocities and wavefunction overlaps are neglected. In the main text, we have explicitly calculated the transmission amplitudes by solving the scattering problem, which is (in our view) the adequate way of treating vdW heterostructures.

By considering the scattering regions from the transmission calculations only, we can also calculate another measure for the ballistic TSP, which is the tunneling density of states (TDOS). The TDOS is defined via the product of the DOS and the velocity of the Bloch bands [3]. Therefore, the TDOS accounts for states that have finite velocity and actually propagate across the Co/hBN/graphene/hBN/Co interface. The TDOS is calculated by employing the tetrahedron method on a $33 \times 33 \times 33$ $k$-mesh, where we postprocess the `Quantum ESPRESSO` output to calculate Bloch band velocities. In fact, based on the spin-resolved DOS, $N_{\uparrow/\downarrow}$, and the Bloch band velocities in $z$-direction, $v_z$, we can calculate three additional measures of the TSP as follows [3]:

$$P_N = \frac{N_\uparrow - N_\downarrow}{N_\uparrow + N_\downarrow} \tag{1}$$

$$P_{Nv} = \frac{\langle Nv_z \rangle_\uparrow - \langle Nv_z \rangle_\downarrow}{\langle Nv_z \rangle_\uparrow + \langle Nv_z \rangle_\downarrow} \tag{2}$$

$$P_{Nv^2} = \frac{\langle Nv_z^2 \rangle_\uparrow - \langle Nv_z^2 \rangle_\downarrow}{\langle Nv_z^2 \rangle_\uparrow + \langle Nv_z^2 \rangle_\downarrow} \tag{3}$$

While $P_N$ is the most natural definition, and $P_{Nv}$ is employed for Andreev reflection in ballistic ferromagnet/superconductor contacts, $P_{Nv^2}$ is the accurate definition of spin polarization of ferromagnets [3]. In Fig.15, we compare the definitions of the TSP with the one from the main text for one exemplary structure. We find that the TSP is very different for the different definitions, but neither of the above nearly captures the qualitative dependence of the transmission calculation results, represented by $P_T$. The reason is, that $P_T$ takes into account the full scattering problem with the vdW interfaces, while the other definitions do not take this into account.

Finally, in Fig. 16, we summarize the spin-resolved transmissions, $T_{\uparrow/\downarrow}$, and the tunneling spin polarization, $P_T$, for six exemplary stacking configurations.

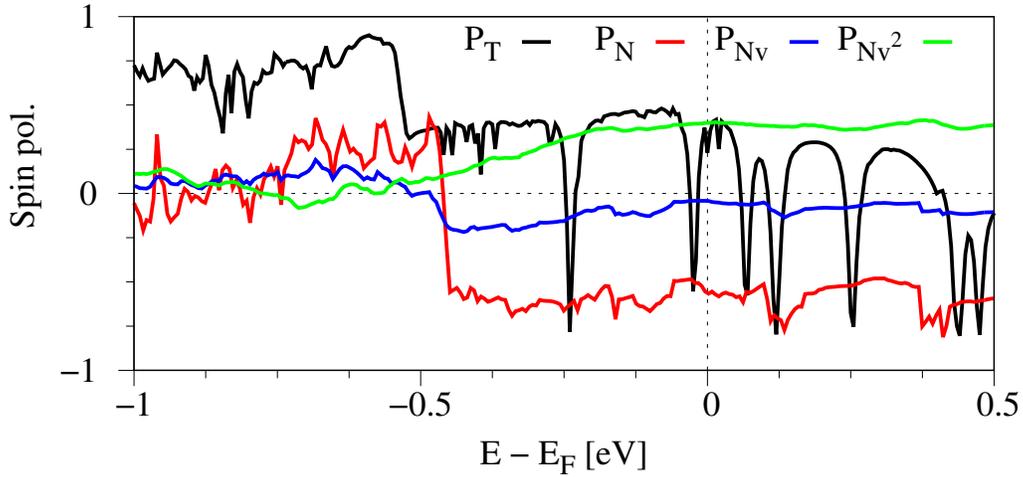

FIG. 15. Comparison of the definitions for the tunneling spin polarization for the Co/hBN/graphene/hBN/Co stack with the $S_{1,2}$ stacking configuration.

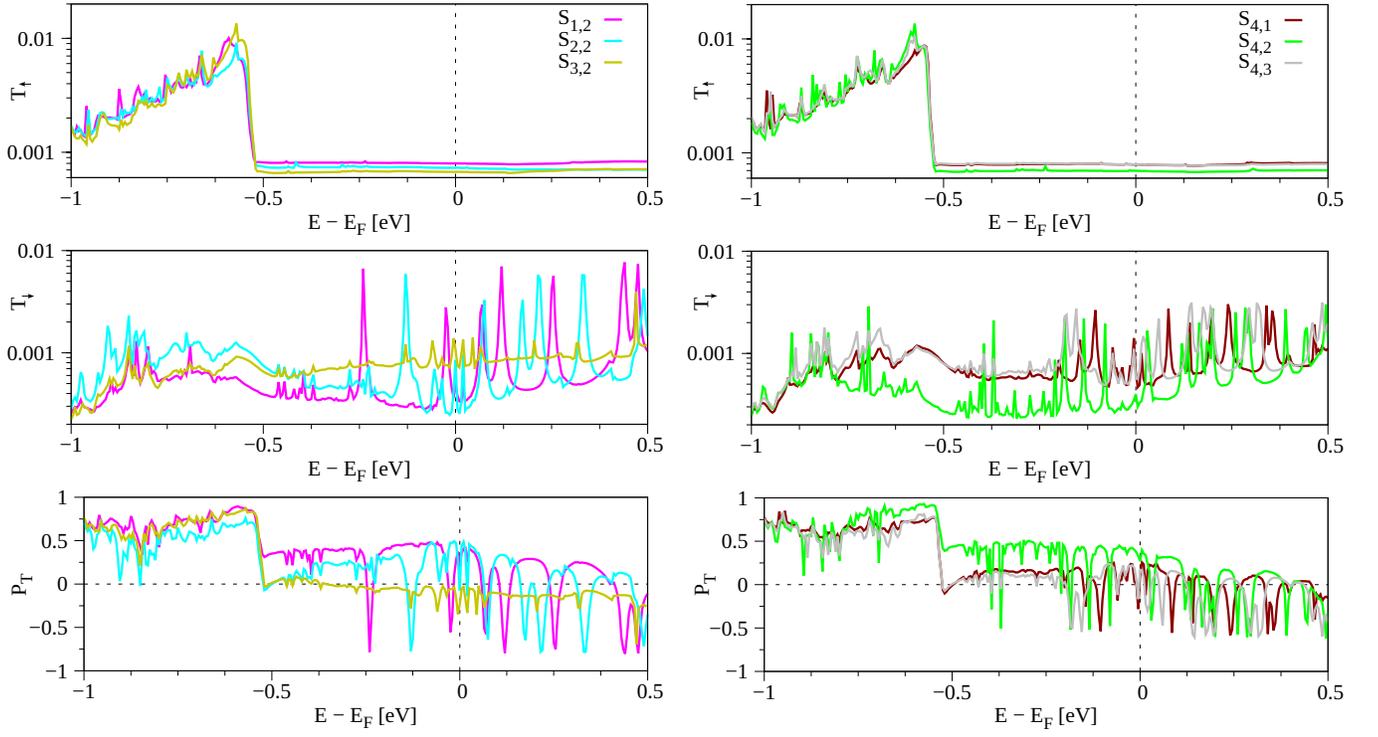

FIG. 16. The spin-resolved transmissions, $T_{\uparrow/\downarrow}$, and the tunneling spin polarization, $P_T$, for six stacking configurations.

## E. The Cobalt Lead

In order to better understand the spin injection from the Co, we also analyze the bulk Co leads. In Fig. 17, we show the band structure and density of states of bulk hcp-Co. From the DOS, it is evident that spin up injection will be quenched once the corresponding DOS sharply drops at around -0.5 eV below the Fermi level. Exactly at this energy, the spin up transmissions, $T_\uparrow$, shown in Fig. 16 for the Co/hBN/graphene/hBN/Co stacks, also sharply drops for all configurations.

We also provide a movie (`spin_and_k_resolved_DOS_hcp_Co.mp4`) on the spin- and k-resolved DOS as function of energy. From there it is evident where in momentum space scattering states can even arise. For tunneling, the wavevector of incoming states needs to be supported by the scattering region, in order to produce a finite transmission. Furthermore, we have calculated the transmission through bulk hcp-Co with the `PWCOND` module of `Quantum Espresso`. We also provide a movie `trans_movie_bulk_Co.mp4` of the spin- and $k$-resolved transmissions as function of energy. In Fig. 18, we compare the definitions of the tunneling spin polarization. The result from the transmission calculation, $P_T$, is very similar to $P_{Nv}$. In other words, the density of states times the Bloch velocities in $z$-direction provide a realistic description of the transport through bulk materials. For multilayer structures, the situation is again different, as we have seen in Fig. 15. There, Brillouin zone spin filtering effects further selectively quench the tunneling modes.

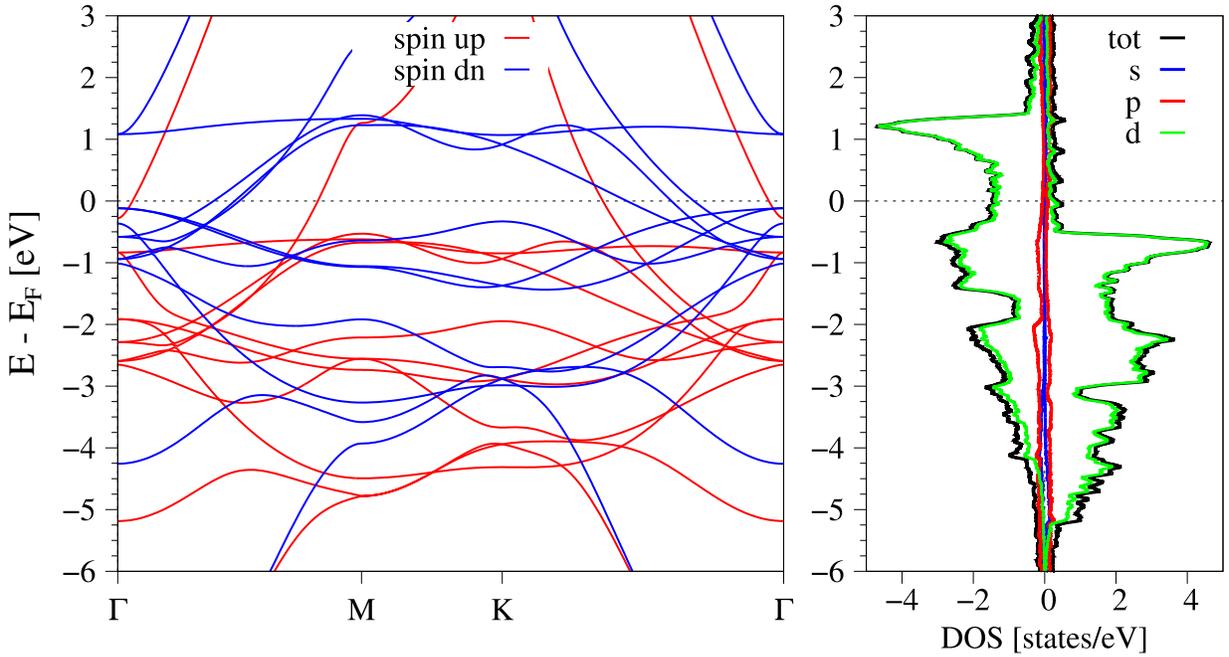

FIG. 17. DFT-calculated band structure of bulk hcp-Co along high-symmetry points and the corresponding spin- and orbital-resolved density of states.

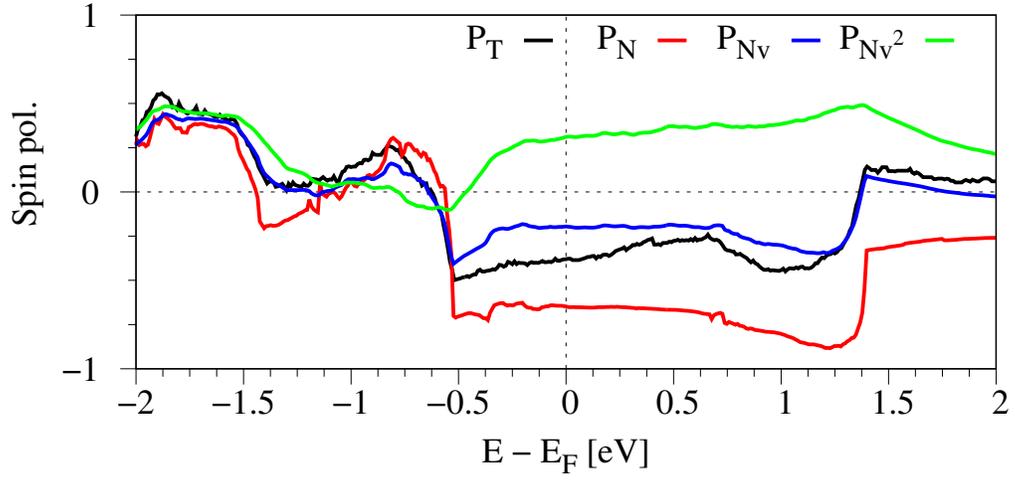

FIG. 18. Comparison of the definitions for the tunneling spin polarization for bulk hcp-Co.

**F. Atomic site overlap**

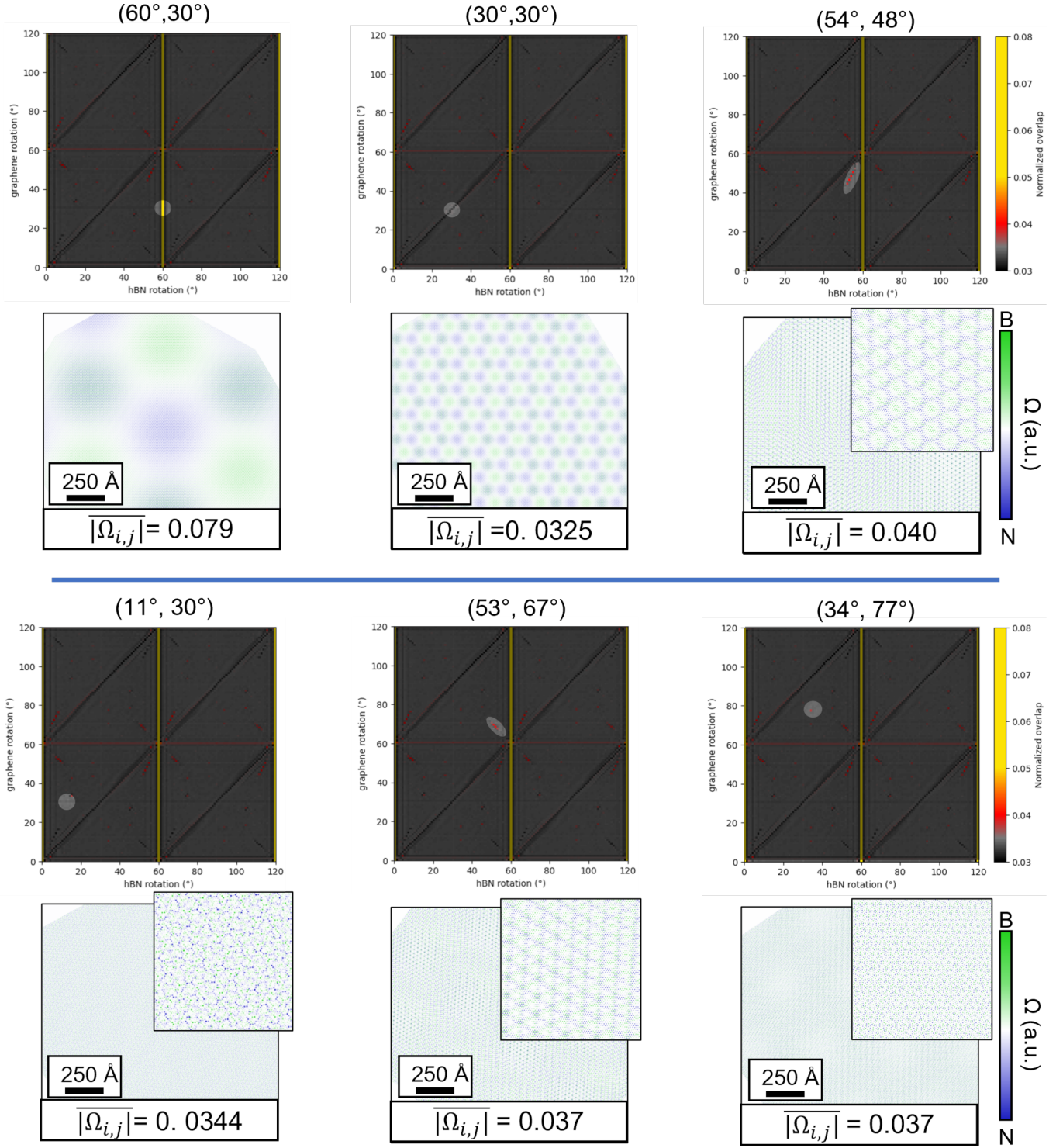

FIG. 19. More twist angle combinations of graphene and hBN. The angles are shown on top of each pair of plots. The top plot shows a masked version of the overlap plot in the main text, highlighting the region inspected. The bottom plot shows the local overlap, where colors indicate vertical alignment with B (N) atoms. Where a pattern is not recognizable, a zoomed in inset is provided.

## II.   EXEMPLARY INPUT FILES

In the following, we provide exemplary input files of the $S_{1,2}$ stack, to reproduce the low energy bands, density of states, and transmission calculations.

──────────── "Bands and DOS Input" ────────────

```
&CONTROL
calculation             = 'scf', ! 'nscf' ! 'bands'
restart_mode            = 'from_scratch',
prefix                  = 'pwscf',
pseudo_dir              = "/potentials",
outdir                  = "./tmp_scf",
/
&SYSTEM
ibrav                   = 0,
nat                     = 7,
ntyp                    = 4,
ecutwfc                 = 75,
ecutrho                 = 800,
occupations             = 'smearing',  ! 'tetrahedra_opt'
vdw_corr                = 'grimme-d2',
degauss                 = 0.00001,
smearing                = 'fd',
noncolin                = .false.,
lspinorb                = .false.,
nspin                   = 2,
starting_magnetization(1) = 0.3,
/
&ELECTRONS
diagonalization         = 'david',
mixing_mode             = 'plain',
electron_maxstep        = 100,
mixing_beta             = 0.7,
conv_thr                = 1e-8,
/
ATOMIC_SPECIES
    Co   58.93319   Co.pbe-n-kjpaw_psl.1.0.0.UPF
    C    12.0107    C.pbe-n-kjpaw_psl.1.0.0.UPF
    B    10.81      B.pbe-n-kjpaw_psl.1.0.0.UPF
    N    14.007     N.pbe-n-kjpaw_psl.1.0.0.UPF
CELL_PARAMETERS angstrom
     2.4903333333333333      0.0000000000000000       0.0000000000000000
    -1.2451666666666666      2.1566919305578470       0.0000000000000000
     0.0000000000000000      0.0000000000000000      29.4000000000000000
ATOMIC_POSITIONS crystal
Co          0.0000000000           0.0000000000           0.3472665524
Co          0.6666666660           0.3333333330           0.4127507306
Co          0.0000000000           0.0000000000           0.4786193768
B           0.6666666660           0.3333333330           0.5459342346
N           0.0000000000           0.0000000000           0.5499943312
C           0.3333333330           0.6666666670           0.6524436072
C           0.6666666660           0.3333333330           0.6524469611
K_POINTS automatic
 300  300   1   0   0   0
! K_POINTS crystal_b
! 3
! 0.3633333      0.3633333      0.00000  200
! 0.3333333      0.3333333      0.00000  200
! 0.3033333      0.3033333      0.00000  1
```

────────────────────────────────────────────

```
&CONTROL
    calculation    = 'scf'
    restart_mode   = 'from_scratch',
    prefix         = 'leadCo',
    wf_collect     = .true.,
    pseudo_dir     = "/potentials",
    outdir         = "./tmp_scf",
/
&SYSTEM
    ibrav          = 4,
    celldm(1)      = 4.705985,
    celldm(3)      = 1.6286,
    nat            = 2,
    ntyp           = 1,
    ecutwfc        = 75,
    ecutrho        = 800,
    occupations    = 'smearing',
    smearing       = 'fd',
    degauss        = 0.001,
    nspin          = 2,
    starting_magnetization(1) =  0.25,
    noncolin       = .false.,
    lspinorb       = .false.,
    nosym          = .true.,
/
&ELECTRONS
    diagonalization = 'david',
    mixing_mode     = 'plain',
    mixing_beta     = 0.2,
    conv_thr        = 1.0d-7,
    electron_maxstep = 100,
/
ATOMIC_SPECIES
Co  58.9332     Co.pbe-n-kjpaw_psl.1.0.0.UPF
ATOMIC_POSITIONS  crystal
 Co     0.00000000000000     0.00000000000000     0.00000000000000     0   0   1
 Co     0.66666666666666     0.33333333333333     0.50000000000000     0   0   1
K_POINTS automatic
36 36 24 0 0 0
```

```
&CONTROL
    calculation     = 'scf'
    restart_mode    = 'from_scratch',
    prefix          = 'scatt',
    pseudo_dir      = "/potentials",
    outdir          = "./tmp_scf",
    verbosity       = 'high'
 /
&SYSTEM
    ibrav           = 4,
    celldm(1)       = 4.705985,
    celldm(3)       = 23.471673957,
    nat             = 22,
    ntyp            = 4,
    ecutwfc         = 75,
    ecutrho         = 800,
    occupations     = 'smearing',
    smearing        = 'fd',
    degauss         = 0.002,
    vdw_corr        = 'grimme-d2',
    nspin           = 2,
    starting_magnetization(1) = 0.25,
    noncolin        = .false.,
    lspinorb        = .false.,
    nosym           = .true.,
 /
&ELECTRONS
    diagonalization = 'david',
    mixing_mode     = 'local-TF',
    mixing_beta     = 0.2,
    conv_thr        = 1.0d-7,
    electron_maxstep = 100,
 /
ATOMIC_SPECIES
Co  58.9332    Co.pbe-n-kjpaw_psl.1.0.0.UPF
C   12.0107    C.pbe-n-kjpaw_psl.1.0.0.UPF
B   10.8110    B.pbe-n-kjpaw_psl.1.0.0.UPF
N   14.0067    N.pbe-n-kjpaw_psl.1.0.0.UPF
ATOMIC_POSITIONS  crystal
Co          0.6666666666        0.3333333333       -0.1008444035    0  0  1
Co          0.0000000000        0.0000000000       -0.0677815744    0  0  1
Co          0.6666666666        0.3333333333       -0.0335669400    0  0  1
Co          0.0000000000        0.0000000000        0.0000000000    0  0  0
Co          0.6666666666        0.3333333333        0.0338436266    0  0  1
Co          0.0000000000        0.0000000000        0.0674994669    0  0  1
Co          0.6666666666        0.3333333333        0.1013737052    0  0  1
Co          0.0000000000        0.0000000000        0.1342410691    0  0  1
B           0.6666666666        0.3333333333        0.1683177873    0  0  1
N           0.0000000000        0.0000000000        0.1702626626    0  0  1
C           0.3333333333        0.6666666666        0.2214993026    0  0  1
C           0.6666666666        0.3333333333        0.2214993214    0  0  1
B           0.6666666666        0.3333333333        0.2746184049    0  0  1
N           0.0000000000        0.0000000000        0.2726855316    0  0  1
Co          0.0000000000        0.0000000000        0.3087096682    0  0  1
Co          0.6666666666        0.3333333333        0.3415706217    0  0  1
Co          0.0000000000        0.0000000000        0.3754494375    0  0  1
Co          0.6666666666        0.3333333333        0.4090901613    0  0  1
Co          0.0000000000        0.0000000000        0.4429336013    0  0  1
Co          0.6666666666        0.3333333333        0.4765040101    0  0  1
Co          0.0000000000        0.0000000000        0.5107288732    0  0  1
Co          0.6666666666        0.3333333333        0.5437953889    0  0  1
K_POINTS automatic
60 60 1 0 0 0
```

```
&INPUTCOND
    outdir     ="./tmp_scf",
    prefixl    = 'leadCo',
    prefixs    = 'scatt',
    prefixr    = 'leadCo',
    tran_file  = 'tm_up.out',   ! 'tm_dn.out'
    tk_plot    = 1,
    ikind      = 1,
    iofspin    = 1,  ! 2
    bds        = 10.3963931,
    energy0    = 0.5d0,
    denergy    = -0.01d0,
    ewind      = 5.0d0,
    epsproj    = 1.d-8,
    delgep     = 1.d-9,
    nz1        = 20,
/
    0
    42 42 0 0
    301
```

```
&control
    calculation = 'scf'
    restart_mode='from_scratch',
    prefix='scatt',
    pseudo_dir="/potentials",
    outdir="./tmp_scf",
    verbosity = 'high'
/
&system
    ibrav= 4,
    celldm(1) = 4.705985,
    celldm(3) = 10.368085016,
    nat= 15,
    ntyp= 4,
    ecutwfc = 75,
    ecutrho = 800,
    occupations='smearing',
    smearing='fd',
    degauss=0.0001,
    vdw_corr = 'grimme-d2',
    nspin = 2,
    starting_magnetization(1) =  0.25,
    noncolin = .false.,
    lspinorb = .false.,
    nosym = .true.,
    noinv = .true.,
    no_t_rev = .true.,
    nosym_evc = .true.,
 /
 &electrons
    diagonalization='david',
    mixing_mode = 'local-TF',
    mixing_beta = 0.2,
    conv_thr =  1.0d-7,
    electron_maxstep = 100,
 /
ATOMIC_SPECIES
Co  58.9332    Co.pbe-n-kjpaw_psl.1.0.0.UPF
C   12.0107    C.pbe-n-kjpaw_psl.1.0.0.UPF
B   10.8110    B.pbe-n-kjpaw_psl.1.0.0.UPF
N   14.0067    N.pbe-n-kjpaw_psl.1.0.0.UPF
ATOMIC_POSITIONS   crystal
Co         0.0000000000         0.0000000000        -0.0005734254
Co         0.6666666666         0.3333333333         0.0758018630
Co         0.0000000000         0.0000000000         0.1518547067
Co         0.6666666666         0.3333333333         0.2284737213
Co         0.0000000000         0.0000000000         0.3026944059
B          0.6666666670         0.3333333330         0.3796848034
N          0.0000000260         0.0000000000         0.3841042260
C          0.3333333330         0.6666666670         0.4993702893
C          0.6666666670         0.3333333330         0.4993704071
B          0.6666666670         0.3333333330         0.6190911754
N          0.0000001976         0.0000000000         0.6146791976
Co         0.0000000000         0.0000000000         0.6961109273
Co         0.6666666666         0.3333333333         0.7703447623
Co         0.0000000000         0.0000000000         0.8469830448
Co         0.6666666666         0.3333333333         0.9230428951
K_POINTS automatic
33 33 33 0 0 0
```

\clearpage

\end{document}